\documentclass{article}

\usepackage[english]{babel}                                                                                           

\usepackage[letterpaper,top=2cm,bottom=2cm,left=3cm,right=3cm,marginparwidth=1.75cm]{geometry}

\usepackage{amsmath}
\usepackage{graphicx}
\usepackage[colorlinks=true, allcolors=blue]{hyperref}
\usepackage{amsthm}
\usepackage{tikz}
\usepackage{centernot}
\usepackage{mathtools}
\usepackage{cite}
\usepackage{ stmaryrd }
\usetikzlibrary{matrix,arrows.meta}
\usepackage{amsmath}
\usepackage{tikz}
\usetikzlibrary{matrix,arrows,decorations.pathmorphing}
\usetikzlibrary{shapes,arrows,positioning}
\usepackage{graphicx}
\usepackage{amsmath}
\usepackage{amssymb}
\usepackage{latexsym}
\usepackage{amssymb,amsmath}
\usepackage{tikz-cd}
\usetikzlibrary{arrows}
\usepackage{amsfonts}
\usepackage{amssymb}
\usepackage{graphicx}
\usepackage{fancyhdr}
\usepackage{fancyvrb}
\usepackage{fancybox}
\usepackage{float}
\usepackage{geometry}
\usepackage{minitoc}
\usepackage{indentfirst}
\usepackage{multicol, color}
\usepackage[outerbars]{changebar}
\usepackage{pifont,textcomp}
\usepackage{amsfonts}
\usepackage{amsmath}
\usepackage{amscd}
\usepackage{array}
\usepackage{multirow}
\usepackage{mathrsfs}
\usepackage{amssymb}
\usepackage{amsthm}
\usepackage{amsmath}
\usepackage{subfig}
\usepackage{graphicx}
\usepackage[colorlinks=true, allcolors=blue]{hyperref}

 \title{Thermodynamic Analysis for Harmonic Oscillator with Position-Dependent Mass}
\author{  Daniel Sabi Takou$^{1,2}$, 	Assimiou Yarou Mora$^{2}$,  and  Gabriel Y. H. Avossevou$^{2}$
 	\space\\\\
 	$^{1}$Ecole Polytechnique d'Abomey Calavi (EPAC-UAC),\\
	Universit\'e d'Abomey-Calavi (UAC), B\'enin\\
	$^{2}$ Unit\'e de Recherche en Physique Th\'eorique (URPT),\\
	Institut de Math\'ematiques et de Sciences  Physiques (IMSP),\\
		01 B.P. 613 Porto-Novo, Rep. du B\'enin\\\\		 
 sabitakoudaniel11@gmail.com$^{1,2}$,  assimiouyaroumora@gmail.com$^{2}$ and\\ gabriel.avossevou@imsp-uac.org$^{2}$ }

\date{}

\begin{document}
\maketitle

\begin{abstract}
In this paper, we examine the thermodynamic behavior of a quantum harmonic oscillator with a position-dependent mass (PDM), where spatial inhomogeneity is modeled through a deformation parameter $\alpha$. Based on the exact energy spectrum, we explore the resulting thermodynamic quantities and superstatistics. Our findings reveal that increasing 
$\alpha$ leads to a decrease in entropy and specific heat, reflecting a confinement-induced reduction in the number of accessible states. The partition function and free energy exhibit smooth behavior across all parameter regimes, indicating the absence of critical phase transitions. This study underscores the influence of mass deformation on quantum thermal responses and demonstrates that, while the overall thermodynamic trends are consistent with those reported in the literature, certain distinctive features emerge due to the specific form of the deformation.
\end{abstract}


\maketitle
\section{Introduction} \label{sec1}

In recent decades, the study of quantum systems with position-dependent mass (PDM) has attracted significant attention due to its wide-ranging applications in condensed matter physics, semiconductor heterostructures, quantum wells, quantum dots, and other nanoscale systems \cite{A11, A12, A13, A14, A15, A16, A17, A18}. These systems play a central role in describing the behavior of charge carriers in non-uniform media, such as He clusters, abrupt heterojunctions, superlattices, and optoelectronic devices. From a theoretical perspective, the PDM Schrödinger equation has been tackled using various analytical methods, including Darboux transformations \cite{A2}, the factorization method \cite{A3}, the Nikiforov–Uvarov (NU) and extended NU methods \cite{A4}, supersymmetric quantum mechanics \cite{A5, A51}, and point canonical transformations \cite{A6}.
calculated the thermodynamic properties of the modified Rosen-Morse potential and examine both the temperature parameter and the maximum quantum state respectively as a function of the 
various thermodynamic properties. 

A variety of potential models such as Kratzer, Pöschl–Teller, Morse, Coulomb, and Hulthén potentials have been investigated under the PDM framework to gain insight into the quantum behavior of these systems \cite{A7, A8}. Recently, significant progress has been made in solving the PDM Schrödinger equation for both confined and unconfined systems, including the infinite square well \cite{A91, A92, A93, A94, A95, A96, A97, A910, A911, A912} and pseudoharmonic oscillators \cite{A10}. Khordad \cite{14} in one of his study examined the 
thermodynamical properties of triangular quantum wires, entropy, specific heat, and internal 
energy. Dong et al. \cite{23} derived exact solutions of the Schrödinger equation involving an exponential-type position-dependent mass. Amir and Iqbal \cite{A910} analytically solved the Schrödinger equation for a particle with PDM confined within an infinite square well (ISW). Inyang et al. \cite{13} calculated the various thermodynamic properties of Eckart
Hellman potential in one of the recent studies. More recently, Iqbal and Rus \cite{25} explored the time evolution of wave packets for a PDM particle in a one-dimensional ISW, highlighting the occurrence of quantum revivals over various time intervals. El Nabulsi \cite{26} introduced a novel formalism to analyze the Schrödinger equation with position-dependent mass, with particular emphasis on applications in semiconductor physics.

Motivated by these developments, we firstly consider in this work a quantum system with a deformed mass profile of the form
\begin{eqnarray}
m(x) = \frac{m_0}{(1 + \alpha x^2)^2},
\end{eqnarray}
where $m_0$ is a constant mass and $\alpha$ is a deformation parameter controlling the spatial variation of the mass. This quadratic deformation generalizes previously studied forms \cite{iqbal2, B1}, and allows for richer physical and mathematical structures. We confine this mass distribution in a one-dimensional harmonic oscillator potential and analyze the corresponding dynamics.

In the second part of this work, we explore the thermodynamic and superstatistical properties of  the ho with Position-Dependent Mass. Specifically, we compute quantities such as the partition function, mean energy, specific heat, Helmholtz free energy, and entropy.

The paper is organized as follows. Section 2 is devoted to the review of the solution methods for the Harmonic Oscillator with Position-Dependent Mass. In Section 3, we evaluated  the thermodynamics and superstatistics properties of model associated with the resulting spectrum. Section 4 focuses on numerical results and discussion. Finally, conclusions are drawn in Section 5.

\section{Review of the Solution Methods for the Harmonic Oscillator with Position-Dependent Mass}\label{sec3}

In this section, presents a comprehensive and self-contained overview of the methodology for deriving the quantum states of a harmonic oscillator with a position-dependent mass \cite{iqbal2, B1}.
We study a quantum harmonic oscillator with a position-dependent mass (PDM), using a Hermitian form of the kinetic energy operator proposed by Mustafa and Mazharimousavi\cite{MOST} and references therein. For a harmonic potential and a mass profile \( m(x) = \frac{m_0}{(1 + \alpha x^2)^2} \), the Schrödinger equation is transformed via a suitable change of variables and wavefunction rescaling.

The most general Hermitian kinetic Hamiltonian for a particle with position-dependent mass (PDM) $m(\hat x)$ is given by \cite{B1}:
\begin{eqnarray}
\hat T=\frac{1}{4}\left[m^\alpha(\hat x)\hat p m^\beta \hat p m^\gamma(\hat x)+ m^\gamma(\hat x)\hat p m^\beta \hat p m^\alpha(\hat x)\right],
\end{eqnarray}
where $\alpha, \beta$ and $\gamma$ satisfy $\alpha + \beta + \gamma = -1$. In this study, we adopt the Mustafa and Mazharimousavi ordering \cite{MOST, B1} with $\alpha = \gamma = -\frac{1}{4}$ and $\beta = -\frac{1}{2}$:
\begin{eqnarray}
\hat T = \frac{1}{2} \frac{1}{m^{1/4}(\hat x)} \hat p \frac{1}{m^{1/2}(\hat x)} \hat p \frac{1}{m^{1/4}(\hat x)}.
\end{eqnarray}

The full Hamiltonian becomes:
\begin{eqnarray}
\hat H = \hat T + V = \frac{1}{2} \frac{1}{m^{1/4}(\hat x)} \hat p \frac{1}{m^{1/2}(\hat x)} \hat p \frac{1}{m^{1/4}(\hat x)} + V(\hat x),
\end{eqnarray}
with $V(x) = \frac{1}{2} m_0 \omega^2 x^2$.

The time-independent Schr"{o}dinger equation then reads:
\begin{eqnarray}\label{a}
E\phi(x) = -\frac{\hbar^2}{2m_0}\sqrt[4]{\frac{m_0}{m(x)}}\frac{d}{dx}\sqrt{\frac{m_0}{m(x)}}\frac{d}{dx}\sqrt[4]{\frac{m_0}{m(x)}} \phi(x) + V(x)\phi(x).
\end{eqnarray}

We consider the mass profile:
\begin{eqnarray}\label{71}
m(x) = \frac{m_0}{(1 + \alpha x^2)^2}, \quad 0 < \alpha < 1.
\end{eqnarray}

Using the transformation $\phi(x) = \sqrt[4]{m(x)/m_0} \psi(x)$, Eq.~\eqref{a} becomes:
\begin{eqnarray}\label{schro}
E\psi(x) = -\frac{\hbar^2}{2m_0}\left[(1 + \alpha x^2) \frac{d}{dx}\right]^2 \psi(x) + \frac{1}{2} m_0 \omega^2 x^2 \psi(x).
\end{eqnarray}

Introducing the variable change:
\begin{eqnarray}
q = \arctan(x\sqrt{\alpha}),
\end{eqnarray}
which maps $x \in (-\infty, \infty)$ to $q \in (-\frac{\pi}{2}, \frac{\pi}{2})$, we obtain:
\begin{eqnarray}
\frac{d^2\psi}{dq^2} + \left(\varepsilon - \kappa^2 \frac{s^2}{c^2}\right)\psi = 0,
\end{eqnarray}
where $\varepsilon = \frac{2m_0E}{\alpha\hbar^2}$, $\kappa = \frac{m_0\omega}{\alpha\hbar}$, $c = \cos q$, $s = \sin q$.

Assuming $\psi(q) = c^\lambda f(s)$, the function $f(s)$ satisfies:
\begin{eqnarray}\label{k}
(1 - s^2)\frac{d^2 f}{ds^2} - (2\lambda + 1)s \frac{d f}{ds} + \left[(\varepsilon - \lambda) - \left(\kappa^2 - \lambda(\lambda - 1)\right)\frac{s^2}{c^2}\right]f = 0.
\end{eqnarray}

To remove the singularity at $c = 0$, we impose:
\begin{eqnarray}
\kappa^2 - \lambda(\lambda - 1) = 0 \quad \Rightarrow \quad \lambda = \frac{1}{2} + \frac{1}{2}\sqrt{1 + 4\kappa^2}.
\end{eqnarray}

This reduces Eq.~\eqref{k} to:
\begin{eqnarray}\label{l}
(1 - s^2)\frac{d^2 f}{ds^2} - (2\lambda + 1)s \frac{d f}{ds} + (\varepsilon - \lambda)f = 0.
\end{eqnarray}

Requiring polynomial solutions (to avoid singularities at $s = \pm 1$) leads to the quantization condition:
\begin{eqnarray}\label{e}
\varepsilon - \lambda = n(n + 2\lambda), \quad n \in \mathbb{N}.
\end{eqnarray}

Substituting back, the final expression for the energy spectrum becomes\cite{B1}:
\begin{eqnarray}\label{en112}
E_n = \hbar\omega \left(n + \frac{1}{2}\right) \sqrt{1 + \frac{\alpha^2\hbar^2}{4m_0^2\omega^2}} + \frac{\alpha\hbar^2}{2m_0} \left(n^2 + 2n + \frac{1}{2}\right).
\end{eqnarray}

\section{Thermodynamics and Superstatistics Properties }

In the thermodynamic and superstatistical analysis of a harmonic oscillator with position-dependent mass, calculating the partition function is a crucial step. It serves as a distribution function in statistical mechanics and provides access to the fundamental properties of the system.

\subsection{Thermodynamics properties}\label{subs2}

\begin{figure}[H]	
 \centering
	\includegraphics[width=7cm, height=6cm]{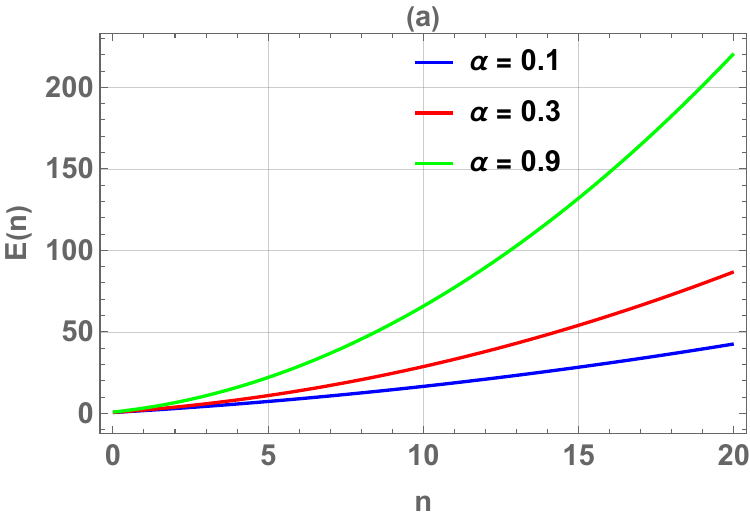}\quad
	\includegraphics[width=7cm, height=6cm]{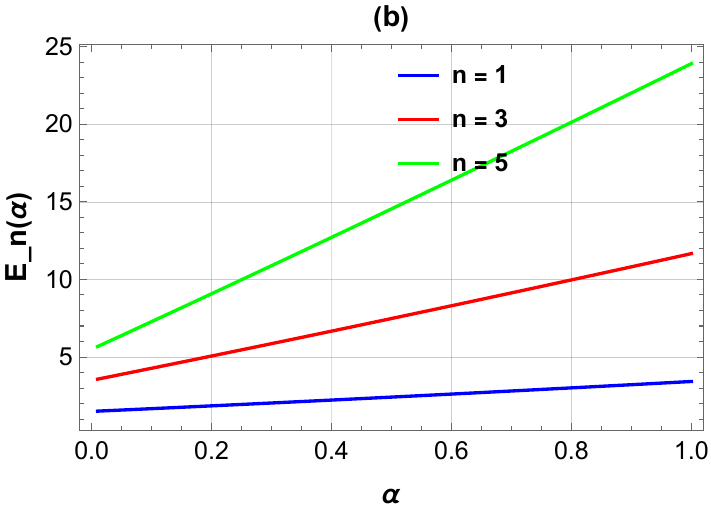}\label{Ea}
\caption{(a). The energy for a PDM harmonic oscillator versus the quantum number $n$ for fix values of the parameter $\alpha$, (b). The energy for a PDM harmonic oscillator versus the the parameter $\alpha$ and the for fix values of quantum number $n$.
}\label{en}
\end{figure}

To evaluate the thermodynamic properties of a harmonic oscillator with position-dependent mass, the energy expression previously derived in Eq.\eqref{en112} is rewritten in a more compact form as follows Eq.\eqref{q122}:
\begin{eqnarray}
    E_n=a\left(n+\frac{1}{2}\right)+b\left(n^2 + 2n + \frac{1}{2}\right)\label{q122}
\end{eqnarray}
where
\begin{equation}
a=\hbar\omega \sqrt{1+\frac{\alpha^2\hbar^2}{4m_0^2\omega^2}},\;\;\;\mbox{and} \quad
	 b=\frac{\alpha\hbar^2}{2m_0\omega}.
\end{equation}
The Energy for a PDM of ho is illustated in Fig.\ref{en}(a) as a function of $n$ and in Fig.\ref{en}(b) as function of $\alpha$. 
The results shows how energy \( E(n) \) evolves with respect to the quantum number \( n \) for three fixed values of \( \alpha \): 0.1, 0.3, and 0.9.
The energy increases nonlinearly with \( n \), and the rate of growth is strongly dependent on the value of \( \alpha \). At higher \( \alpha \), the energy levels for a given \( n \) are significantly higher, indicating that the system becomes more energetically demanding as \( \alpha \) increases.
The Fig.\ref{en}(b) show that energy increases monotonically with \( \alpha \) for each fixed value of \( n \). Additionally, higher energy states (\( n = 3 \) and \( n = 5 \)) remain consistently above lower ones, indicating that energy grows with \( n \) as well. 
 Fig.\ref{en}(b) illustrates the variation of the energy levels \( E_n(\alpha) \) as a function of the parameter \( \alpha \), for three quantum states: \( n = 1 \), \( n = 3 \), and \( n = 5 \).

This behavior suggests that \( \alpha \) acts as a deformation or coupling parameter that enhances the confining potential or interaction strength, resulting in steeper energy scaling with the quantum number \( n \).

The curves exhibit a convex profile, suggesting that the rate of increase in energy becomes more pronounced as \( \alpha \) increases. This trend may reflect a growing confinement, curvature, or interaction strength in the system governed by \( \alpha \).

 Once the energy spectrum is expressed in a simplified form, one can readily proceed to the evaluation of the partition function, which serves as a basis for deriving the associated thermodynamic properties. For systems with bound states, the partition function is generally defined as follows equation \cite{ekonnn, ek1, ek2, ek3, ek4, ek5}.

 \begin{eqnarray}\label{parti}
     Z(\beta)=\sum_{n=0}^{\infty} e^{-\beta E_{n}}, \;\;\;\beta=\frac{1}{k_B T}
 \end{eqnarray}
 Where $k_B$ is the Boltzmann constant and $E_n$ is energy of the bound state. By inserting the expressions of the energy spectrum derived in Eq. \eqref{q122} into Eq. \eqref{parti}, and using Mathematica 13.3 software, we obtain the $\alpha$-deformed partition function within the framework of superstatistics for the position-dependent mass harmonic oscillator, in the  presence of the deformation parameter $\alpha$.
 
\begin{eqnarray}\label{parti1}
Z(\beta) = \frac{e^{\frac{(a^2+2ab+2b^2)}{4b} \beta} \sqrt{\pi}}{2\sqrt{b\beta}} 
\left(
-\operatorname{Erf} \left[ \frac{1}{2} (a+2b) \sqrt{\frac{\beta}{b}} \right] 
+ \operatorname{Erf} \left[ \frac{1}{2} (a+4b) \sqrt{\frac{\beta}{b}} \right] 
\right)
\end{eqnarray}
After a careful computation of the partition function associated with the model eq. \eqref{parti1}, we now proceed to extract the corresponding thermodynamic quantities. Using the compact expression of the partition function \( Z(\beta) \),  the mean internal energy \( U(\beta) \), the  heat specific \( C(\beta) \), the entropy \( S(\beta) \) and the thermodynamic potentials such as the Helmholtz free energy \( F(\beta) \),  can be obtained from the standard thermodynamic identities \cite{ekonnn, ek1, ek2, ek3, ek4, ek5}:

\begin{eqnarray}
&U(\beta)& =-\frac{\partial(\ln Z(\beta))}{\partial \beta}\\
&=& \frac{e^{-3 \beta - \frac{a^2 \beta}{2b} - 5b \beta} \left( 2 \sqrt{b \beta} \, e^{\frac{(a^2+2b)^2 \beta}{4b}} \left( 2 b (-2 + e^{(a+3b) \beta}) + a (-1 + e^{(a+3b) \beta}) \right) \right.}{4b \beta \sqrt{\pi} \left( -\operatorname{Erf} \left[ \frac{1}{2} (a + 2b) \sqrt{\frac{\beta}{b}} \right]+ \operatorname{Erf} \left[ \frac{1}{2} (a + 4b) \sqrt{\frac{\beta}{b}} \right] \right)} \nonumber\\
&+& \frac{ \left. (a^2 \beta + 2 b^2 \beta + 2 b (-1 + a \beta)) \, e^{3a \beta + \frac{a^2 \beta}{2b} + 5b \beta} \sqrt{\pi} \left( \operatorname{Erf} \left[ \frac{1}{2} (a + 2b) \sqrt{\frac{\beta}{b}} \right] - \operatorname{Erf} \left[ \frac{1}{2} (a + 4b) \sqrt{\frac{\beta}{b}} \right] \right) \right)}{4b \beta \sqrt{\pi} \left( -\operatorname{Erf} \left[ \frac{1}{2} (a + 2b) \sqrt{\frac{\beta}{b}} \right]+ \operatorname{Erf} \left[ \frac{1}{2} (a + 4b) \sqrt{\frac{\beta}{b}} \right] \right)}\nonumber\\
\end{eqnarray}

\begin{eqnarray}
	C(\beta) &=&k_B\beta^2\frac{\partial^2}{\partial \beta^2}\ln Z(\beta)\\
    &=& \Bigg( \beta e^{-5a\beta - \frac{3a^2 \beta}{4b} - 9 b \beta} \bigg[
	2 b \beta \sqrt{\frac{\beta}{b}} e^{\frac{(a+2b)^2 \beta}{4b}} 
	\left( a - a e^{(a+3b) \beta} - 2b (-2 + e^{(a+3b) \beta}) \right)^2 \nonumber \\
	& - &4 b \sqrt{b \beta} e^{5a\beta + \frac{3a^2 \beta}{4b} + 9 b \beta} 
	\pi \operatorname{Erf} \left[ \frac{1}{2} (a + 2b) \sqrt{\frac{\beta}{b}} \right]^2 
	- \beta e^{3 a \beta + \frac{a^2 \beta}{2b} + 5 b \beta} \nonumber \\
	& \times& \bigg( 6 a^2 b \beta (-2 + e^{(a+3b) \beta}) + a^3 \beta (-1 + e^{(a+3b) \beta}) 
	+ 4 b^2 \bigg(-2 + e^{(a+3b) \beta} + 2 b \beta \bigg(-8   \nonumber \\
	&+& e^{(a+3b) \beta}\bigg)\bigg)  + 2 a b (-1 + e^{(a+3b) \beta} + 6 b \beta (-4 + e^{(a+3b) \beta})) \bigg) 
	\sqrt{\pi} \operatorname{Erf} \left[ \frac{1}{2} (a + 4b) \sqrt{\frac{\beta}{b}} \right] 
	 \nonumber \\
	&-& 4 b \sqrt{b \beta} e^{5 a \beta + \frac{3a^2 \beta}{4b} + 9 b \beta} \pi \operatorname{Erf} \left[ \frac{1}{2} (a + 4b) \sqrt{\frac{\beta}{b}} \right]^2 
	+ \sqrt{\pi} \operatorname{Erf} \left[ \frac{1}{2} (a + 2b) \sqrt{\frac{\beta}{b}} \right] 
	  \nonumber \\
	& \times& \beta e^{3 a \beta + \frac{a^2 \beta}{2b} + 5 b \beta} \bigg( 6 a^2 b \beta (-2 + e^{(a+3b) \beta}) + a^3 \beta (-1 + e^{(a+3b) \beta}) 
	+ 4 b^2 \bigg(-2 + e^{(a+3b) \beta}   \nonumber \\
	&+& 2 b \beta (-8 + e^{(a+3b) \beta})\bigg)  + 2 a b (-1 + e^{(a+3b) \beta}) + 6 b \beta (-4 + e^{(a+3b) \beta}) \bigg) \nonumber \\
	&+& 8 b \sqrt{b \beta} e^{5 a \beta + \frac{3a^2 \beta}{4b} + 9 b \beta} 
	\sqrt{\pi} \operatorname{Erf} \left[ \frac{1}{2} (a + 4b) \sqrt{\frac{\beta}{b}} \right] \Bigg) \nonumber \\
	&& \Bigg/ \left( 8 (b \beta)^{3/2} \pi 
	\left( \operatorname{Erf} \left[ \frac{1}{2} (a + 2b) \sqrt{\frac{\beta}{b}} \right] 
	- \operatorname{Erf} \left[] \frac{1}{2} (a + 4b) \sqrt{\frac{\beta}{b}} \right] \right)^2 \right)
\end{eqnarray}

\begin{eqnarray}
	S(\beta)&=& k_B\ln Z(\beta)-k_B \beta \frac{\partial(\ln Z(\beta))}{\partial \beta}\\
    &=&\bigg(\sqrt{\frac{\beta}{b}} e^{-3 a \beta - \frac{a^2 \beta}{2 b} - 
		5 b \beta} \bigg(2 \beta e^{\frac{(a + 2 b)^2 \beta}{4 b}} (a - a e^{(a + 3 b) \beta} - 
	2 b (-2 + e^{(a + 3 b) \beta})\bigg) \nonumber\\
	&-&\sqrt{\frac{\beta}{b}} (a^2 \beta + 2 b^2 \beta + 2 b (-1 + a \beta)) e^{3 a \beta + \frac{a^2 \beta}{2 b} + 5 b \beta} \sqrt{\pi}
	\operatorname{Erf} \left[ \frac{1}{2} (a + 2b) \sqrt{\frac{\beta}{b}} \right] \nonumber \\
	& +& \sqrt{\frac{\beta}{b}} \bigg(a^2 \beta + 2 b^2 \beta + 2 b (-1 + a \beta)\bigg) e^{3 a \beta + \frac{a^2 \beta}{2 b} + 5 b \beta} \sqrt{\pi}
	\operatorname{Erf} \left[ \frac{1}{2} (a + 4b) \sqrt{\frac{\beta}{b}} \right] \bigg)\bigg)\bigg/ \nonumber \\
	&&\bigg(4\beta\pi\bigg(\operatorname{Erf} \left[ \frac{1}{2} (a + 2b) \sqrt{\frac{\beta}{b}} \right]-\operatorname{Erf} \left[ \frac{1}{2} (a + 4b) \sqrt{\frac{\beta}{b}} \right]\bigg)\bigg)\nonumber\\
	&& + \log \left(\frac{e^{\frac{(a^2 + 2 a b + 2 b^2) \beta}{4 b}}
		\sqrt{\pi} \left(-\operatorname{Erf} \left[ \frac{1}{2} (a + 2b) \sqrt{\frac{\beta}{b}} \right] + 
		\operatorname{Erf} \left[ \frac{1}{2} (a + 4b) \sqrt{\frac{\beta}{b}} \right] \right)}{2 \sqrt{b \beta}} \right).
\end{eqnarray}

\begin{eqnarray}
 F(\beta)&=&-\frac{1}{\beta}\ln Z(\beta)\\
&=&-\frac{1}{\beta} \log \left( \frac{e^{\frac{(a^2 + 2 a b + 2 b^2) \beta}{4 b}}
	\sqrt{\pi} \left(-\operatorname{Erf} \left[ \frac{1}{2} (a + 2b) \sqrt{\frac{\beta}{b}} \right] 
	+ \operatorname{Erf} \left[ \frac{1}{2} (a + 4b) \sqrt{\frac{\beta}{b}} \right] \right)}{2 \sqrt{b \beta}} \right)
\end{eqnarray}

\subsection{Superstatistics Properties }

Superstatistics is a statistical framework developed to describe driven, non-equilibrium systems characterized by fluctuations in intensive parameters, such as the inverse temperature \( \beta \), chemical potential, or energy\cite{sth1, sth2, sth3, sth4}. These fluctuations occur over spatiotemporal scales and are typically captured by extending the conventional Boltzmann factor into a more general form known as the \textit{effective Boltzmann factor} \cite{sth4, sth5, sth6, sth7}.

In this context, superstatistics can be viewed as a superposition of different local equilibrium statistics. The standard approach involves taking the Laplace transform of the probability density function \( f(\beta') \), which results in the generalized Boltzmann factor \cite{sth5, sth7}:

\begin{equation}
    B_E(\beta) = \int_0^\infty e^{-\beta' E} f(\beta', \beta) \, d\beta'.
\end{equation}

When \( f(\beta', \beta) \) is modeled by a Dirac delta function \( \delta(\beta - \beta') \), the integral simplifies, and a deformation parameter \( q \) can be introduced to yield the generalized Boltzmann factor:

\begin{equation}
    B_E^{(q)}(\beta) = e^{-\beta E} \left(1 + \frac{q}{2} \beta^2 E^2 \right).
\end{equation}

Here, \( q \in [0, 1] \) is a deformation parameter that quantifies the departure from classical Boltzmann-Gibbs statistics. In the limit \( q \rightarrow 0 \), standard statistical mechanics is recovered.

The partition function in the superstatistical framework is then defined as:

\begin{equation}
    Z_s = \int_0^\infty B_E^{(q)}(\beta) \, dn\label{sthe1}.
\end{equation}

For discrete systems, such as quantum harmonic oscillators, this can be written as a summation over energy levels:

\begin{equation}
    Z = \sum_{n=0}^\infty B_E(n\beta).
\end{equation}

At high temperature limits (\( T \gg 1 \)), the discrete summation may be approximated by an integral to simplify analysis.

When applied to specific systems like the harmonic oscillator with position-dependent mass, this formalism enables the derivation of modified thermodynamic quantities. These generalized functions are valid for all values of \( q \) and explicitly depend on the system’s energy spectrum. Thus, superstatistics provides a robust framework for analyzing complex, non-linear, and far-from-equilibrium systems where traditional statistical mechanics falls short.\\
Using Eq.\eqref{sthe1}and Mathematica 13.3, the superstatistics partition function equation is given as
\begin{eqnarray}
Z_s(a,b,\beta,q)&=&
\frac{1}{64 b^{5/2} \sqrt{\beta}} e^{-\frac{1}{2}(a+b)\beta} \bigg(\sqrt{\beta}
\bigg(12ab^{3/2} + 24b^{5/2} - 2a^3 \sqrt{b} \beta + a^4 \beta^{3/2} \nonumber\\
&\times &e^{\frac{(a+2b)^2 \beta}{4b}}\sqrt{\pi}\bigg) q - 4a^2 b \beta\bigg( \sqrt{b \beta} + (1 - a \beta) 
e^{\frac{(a+2b)^2 \beta}{4b}} \sqrt{\pi}\bigg) q + 4b^4 \beta^2 e^{\frac{(a+2b)^2 \beta}{4b}} \sqrt{\pi} q\nonumber\\
& + &8b^3 \beta (-1 + a \beta) e^{\frac{(a+2b)^2 \beta}{4b}} \sqrt{\pi} q + 4b^2 e^{\frac{(a+2b)^2 \beta}{4b}} \sqrt{\pi}
\bigg(8 + (3 - 2a \beta + 2a^2 \beta^2) q \bigg)\nonumber\\
&-& e^{\frac{(a+2b)^2 \beta}{4b}} \sqrt{\pi} \bigg(
a^4 \beta^2 q + 4b^4 \beta^2q+4a^2b\beta (-1 + a \beta) q + 8b^3 \beta (-1 + a \beta) q\nonumber\\ 
&+& 4b^2 
\bigg(8 + (3 - 2a \beta + 2a^2 \beta^2) q \bigg)
\mathrm{Erf} \bigg[ \frac{(a+2b)\beta}{2 \sqrt{b \beta}} \bigg]
\bigg)
\bigg)\label{sthe2}.
\end{eqnarray}

Using the Superstatistic  partition function in Eq.\eqref{sthe2}, the  Superstatistics   thermodynamics properties are obtained as follows.

\begin{enumerate}
    \item   Vibrational mean energy 
 
\begin{eqnarray}
 U_s(a,b,\beta,q)&=&-\frac{\partial}{\partial \beta}\ln Z_s(\beta)\\
 &=&-\bigg[4a^3 \bigg( -2b^{3/2}\beta^{3/2}\sqrt{b\beta} + b^{5/2}\beta^{5/2} \sqrt{b\beta} + 2b^2\beta^2 \left(1 + \sqrt{b\beta}e^\frac{(a + 2b)^2 \beta}{4b} \sqrt{\pi} \right) \nonumber\\
&&  + 2b^3 \beta^3 \left(-3 + 4\sqrt{b\beta}e^\frac{(a + 2b)^2 \beta}{4b} \sqrt{\pi} \right)\bigg) q  \nonumber\\
&& + 2a^5 b \beta^3 \left(-1 + 3\sqrt{b\beta}e^\frac{(a + 2b)^2 \beta}{4b} \sqrt{\pi} \right)q + 2a^4 b \beta^2 \bigg( -4b\beta+\sqrt{b\beta}e^\frac{(a + 2b)^2 \beta}{4b} \sqrt{\pi}\nonumber \\
&& + 9b^{3/2}\beta^{3/2}e^\frac{(a + 2b)^2 \beta}{4b} \sqrt{\pi}\bigg) q+ a^{6}\beta^{3}\sqrt{b\beta}e^\frac{(a + 2b)^2 \beta}{4b}\sqrt{\pi}q+8ab^{2}\beta\bigg(-8\nonumber\\
&&-\bigg(3+3b\beta+5b^{2}\beta^2\bigg)q+\sqrt{b\beta}e^\frac{(a + 2b)^2 \beta}{4b}\sqrt{\pi}\bigg(8+q+2b\beta q+3b^2\beta^2q\bigg)\bigg)\nonumber\\
&&+4a^2b^2\beta\bigg(-2b\beta q-10b^{2}\beta^2q+\sqrt{b\beta}e^\frac{(a + 2b)^2 \beta}{4b}\sqrt{\pi}(8+q+4b\beta q+9b^2\beta^2q)\bigg)\nonumber\\
&&+8b^3\bigg(-6b^{3/2}\beta^{3/2}\sqrt{b\beta}q+b^3\beta^3\bigg(-2+\sqrt{b\beta}e^\frac{(a + 2b)^2 \beta}{4b}\sqrt{\pi}\bigg)q\nonumber\\
&&+b^2\beta^2\bigg(4+\sqrt{b\beta}e^\frac{(a + 2b)^2 \beta}{4b}\sqrt{\pi}\bigg)q-\sqrt{b\beta}e^\frac{(a + 2b)^2 \beta}{4b}\sqrt{\pi}(8+3q)\nonumber\\
&&+b\beta\bigg(\sqrt{b\beta}e^\frac{(a + 2b)^2 \beta}{4b}\sqrt{\pi}(8+q)-2(8+3q)\bigg)\bigg)-\sqrt{b\beta}\bigg(a^2\beta+2b^2\beta\nonumber\\
&&+2b(-1+a\beta)\bigg)e^\frac{(a + 2b)^2 \beta}{4b}\sqrt{\pi}\bigg( a^4 \beta^2 q + 4 b^4 \beta^2 q + 4 a^2 b \beta (1 + a \beta) q  \nonumber\\
&&  + 8 b^3 \beta (1 + a \beta) q + 4 b^2 \bigg( 8 + (3 + 2 a \beta + 2 a^2 \beta^2) q \bigg) \bigg) 
\text{Erf} \left[\frac{(a+2b)\beta}{2\sqrt{b\beta}} \right] 
\bigg] \bigg/\nonumber
\end{eqnarray}
\begin{eqnarray}
&&\bigg[4b^{3/2}\beta^{3/2}\bigg(12ab^{3/2}\sqrt{\beta}q+	24b^{5/2}\sqrt{\beta}q-2a^3\sqrt{b}\beta^{3/2}q-4a^2b\beta\bigg(\sqrt{b\beta}\nonumber\\
&&+(1-a\beta)e^\frac{(a + 2b)^2 \beta}{4b}\sqrt{\pi}\bigg)q+a^4\beta^2e^\frac{(a + 2b)^2 \beta}{4b}\sqrt{\pi}q+4b^4\beta^2e^\frac{(a + 2b)^2 \beta}{4b}\sqrt{\pi}q\nonumber\\
&&+8b^2\beta(-1+a\beta)e^\frac{(a + 2b)^2 \beta}{4b}\sqrt{\pi}q+4b^2e^\frac{(a + 2b)^2 \beta}{4b}\sqrt{\pi}\bigg( 8 \nonumber\\
&&+ (3 - 2 a \beta + 2 a^2 \beta^2) q \bigg)-e^\frac{(a + 2b)^2 \beta}{4b}\sqrt{\pi}\bigg(a^4 \beta^2 q +4b^4 \beta^2 q \nonumber\\
&&+ 4 a^2 b \beta (-1 + a \beta) q + 8 b^3 \beta (-1 + a \beta)q\nonumber\\
&&+4b^2\bigg( 8 + (3 - 2 a \beta + 2 a^2 \beta^2) q \bigg)\bigg)\text{Erf} \left[\frac{(a+2b)\beta}{2\sqrt{b\beta}} \right] \bigg)\bigg].
\end{eqnarray}

\item  Free energy  
\begin{eqnarray}    
F_s(a,b,\beta,q)&=&-\frac{1}{\beta}\ln{Z_s(\beta)}\nonumber\\
&=&-\frac{1}{\beta}\times \ln\bigg( 
  \frac{1}{64 b^{5/2} \sqrt{\beta}} e^{-\frac{1}{2}(a+b)\beta} \bigg(\sqrt{\beta}
\bigg(12ab^{3/2} + 24b^{5/2} + 2a^3 \sqrt{b} \beta + a^4 \beta^{3/2}  \nonumber\\
&\times &   e^{\frac{(a+2b)^2 \beta}{4b}}\sqrt{\pi}\bigg) q - 4a^2 b \beta\bigg( \sqrt{b \beta} + (1 - a \beta) 
e^{\frac{(a+2b)^2 \beta}{4b}} \sqrt{\pi}\bigg) q + 4b^4 \beta^2 e^{\frac{(a+2b)^2 \beta}{4b}} \sqrt{\pi} q \nonumber\\
&  + & 8b^3 \beta (-1 + a \beta) e^{\frac{(a+2b)^2 \beta}{4b}} \sqrt{\pi} q + 4b^2 e^{\frac{(a+2b)^2 \beta}{4b}} \sqrt{\pi}
\bigg(8 + (3 - 2a \beta + 2a^2 \beta^2) q \bigg) \nonumber\\
&-& e^{\frac{(a+2b)^2 \beta}{4b}} \sqrt{\pi} \bigg(
a^4 \beta^2 q + 4b^4 \beta^2q+4a^2b\beta (-1 + a \beta) q + 8b^3 \beta (-1 + a \beta) q \\
& + & 4b^2 
\bigg(8 + (3 - 2a \beta + 2a^2 \beta^2) q \bigg)
\mathrm{Erf} \bigg[ \frac{(a+2b)\beta}{2 \sqrt{b \beta}} \bigg]
\bigg)\nonumber
\bigg)\bigg).
\end{eqnarray}  
\item Entropy   
\begin{eqnarray}
S_s(a,b,\beta,q)&=&k_B \Bigg( - \beta \Bigg(
	 4 a^3 \bigg( -2 b^{3/2} \beta^{3/2} \sqrt{b \beta} + b^{5/2} \beta^{5/2} \sqrt{b \beta} + 2 b^2 \beta^2 \bigg(1  \nonumber\\
	& + & \sqrt{b \beta} e^{\frac{(a + 2b)^2 \beta}{4b}} \sqrt{\pi} \bigg) + 2 b^3 \beta^3 \left(-3 + 4 \sqrt{b \beta} e^{\frac{(a + 2b)^2 \beta}{4b}} \sqrt{\pi} \right) \bigg) q \nonumber\\
	&+& 2 a^5 b \beta^3 \left(-1 + 3 \sqrt{b \beta} e^{\frac{(a + 2b)^2 \beta}{4b}} \sqrt{\pi} \right) q \nonumber\\
	&+& 2 a^4 b \beta^2 \left( -6 b \beta + b\beta + \sqrt{b \beta} e^{\frac{(a + 2b)^2 \beta}{4b}} \sqrt{\pi} + 9 (b \beta)^{3/2} e^{\frac{(a + 2b)^2 \beta}{4b}} \sqrt{\pi} \right) q \nonumber\\
	&+& a^6 \beta^3 \sqrt{b \beta} e^{\frac{(a + 2b)^2 \beta}{4b}} \sqrt{\pi} q + 8 a b^3 \beta \bigg( -8 - (3 +3 b \beta + 5 b^2 \beta^2) q \nonumber\\
	& + & \sqrt{b \beta} e^{\frac{(a + 2b)^2 \beta}{4b}} \sqrt{\pi} (8 + q + 2 b \beta q + 3 b^2 \beta^2 q) \bigg) \nonumber\\
	&+& 4 a^2 b^2 \beta \bigg( - 2 b \beta q - 10 b^2 \beta^2 q  \nonumber\\
	& + &  \sqrt{b \beta} e^{\frac{(a + 2b)^2 \beta}{4b}} \sqrt{\pi} (8 + q + 4 b \beta q + 9 b^2 \beta^2 q) \bigg) \nonumber\\
	&+& 8 b^3 \bigg( -6 b^{3/2} \beta^{3/2} \sqrt{b \beta} q + b^3 \beta^3 \left( -2 + \sqrt{b \beta} e^{\frac{(a + 2b)^2 \beta}{4b}} \sqrt{\pi} \right) q \nonumber\\
	&+ &  b^2 \beta^2 \left( 4 + \sqrt{b \beta} e^{\frac{(a + 2b)^2 \beta}{4b}} \sqrt{\pi} \right) q - \sqrt{b \beta} e^{\frac{(a + 2b)^2 \beta}{4b}} \sqrt{\pi} (8 + 3q) \nonumber
\end{eqnarray}
\begin{eqnarray}
	& + &  b \beta \left( \sqrt{b \beta} e^{\frac{(a + 2b)^2 \beta}{4b}} \sqrt{\pi} (8 + q) - 2 (8 + 3q) \right) \bigg) \nonumber\\
	&-& \sqrt{b \beta} (a^2 \beta + 2 b^2 \beta + 2 b (-1 + a \beta)) e^{\frac{(a + 2b)^2 \beta}{4b}} \sqrt{\pi} \nonumber\\
	& & \times \bigg( a^4 \beta^2 q + 4 b^4 \beta^2 q + 4 a^2 b \beta (1 + a \beta) q + 8 b^3 \beta (1 + a \beta) q + 4 b^2 (8  \nonumber\\
	&+ &  (3 + 2 a \beta + 2 a^2 \beta^2) q) \bigg) \text{Erf} \left( \frac{(a + 2b) \beta}{2 \sqrt{b \beta}} \right)
	\Bigg) \Bigg/ \Bigg( 4 (b \beta)^{3/2} \bigg(\nonumber\\
	& & 12 a b^{3/2} \sqrt{\beta} q + 24 b^{5/2} \sqrt{\beta} q - 2 a^3 \sqrt{b} \beta^{3/2} q \nonumber\\
	&-& 4 a^2 b \beta \left( \sqrt{b \beta} + (1 - a \beta) e^{\frac{(a + 2b)^2 \beta}{4b}} \sqrt{\pi} \right) q \nonumber\\
	&+& a^4 \beta^2 e^{\frac{(a + 2b)^2 \beta}{4b}} \sqrt{\pi} q + 4 b^4 \beta^2 e^{\frac{(a + 2b)^2 \beta}{4b}} \sqrt{\pi} q \nonumber\\
	&+& 8 b^3 \beta (-1 + a \beta) e^{\frac{(a + 2b)^2 \beta}{4b}} \sqrt{\pi} q \nonumber\\
	&+& 4 b^2 e^{\frac{(a + 2b)^2 \beta}{4b}} \sqrt{\pi} (8 + (3 - 2 a \beta + 2 a^2 \beta^2) q) \nonumber\\
	&-& e^{\frac{(a + 2b)^2 \beta}{4b}} \sqrt{\pi} \bigg( a^4 \beta^2 q + 4 b^4 \beta^2 q + 4 a^2 b \beta (-1 + a \beta) q \nonumber\\
	& +&  8 b^3 \beta (-1 + a \beta) q + 4 b^2 (8 + (3 - 2 a \beta + 2 a^2 \beta^2) q) \bigg) \nonumber\\
	& & \times \text{Erf} \left[ \frac{(a + 2b) \beta}{2 \sqrt{b \beta}} \right]
	\bigg) \Bigg) + \ln \Bigg[ \frac{1}{64 b^{5/2} \sqrt{\beta}} e^{-\frac{1}{2} (a + b) \beta} \nonumber\\
	&\times &\Bigg( \sqrt{\beta} \left(12 a b^{3/2} + 24 b^{5/2} - 2 a^3 \sqrt{b} \beta + a^4 \beta^{3/2} e^{\frac{(a + 2b)^2 \beta}{4b}} \sqrt{\pi} \right) q \nonumber \\
	&-& 4 a^2 b \beta \left( \sqrt{b \beta} + (1 - a \beta) e^{\frac{(a + 2b)^2 \beta}{4b}} \sqrt{\pi} \right) q \nonumber\\
	&+& 4 b^4 \beta^2 e^{\frac{(a + 2b)^2 \beta}{4b}} \sqrt{\pi} q + 8 b^3 \beta (-1 + a \beta) e^{\frac{(a + 2b)^2 \beta}{4b}} \sqrt{\pi} q \nonumber\\
	&+& 4 b^2 e^{\frac{(a + 2b)^2 \beta}{4b}} \sqrt{\pi} (8 + (3 - 2 a \beta + 2 a^2 \beta^2) q) \nonumber\\
	&-& e^{\frac{(a + 2b)^2 \beta}{4b}} \sqrt{\pi} \bigg( a^4 \beta^2 q + 4 b^4 \beta^2 q + 4 a^2 b \beta (-1 + a \beta) q \nonumber\\
	&  + & 8 b^3 \beta (-1 + a \beta) q + 4 b^2 (8 + (3 - 2 a \beta + 2 a^2 \beta^2) q) \bigg)  \text{Erf} \left[ \frac{(a + 2b) \beta}{2 \sqrt{b \beta}} \right]
	\Bigg) \Bigg]
\end{eqnarray}

    \item  Specific heat capacity 
    
 \begin{eqnarray} C_s(a,b,\beta,q)=k_B\beta^2\frac{\partial^2}{\partial \beta^2}\ln Z_s(a,b,\beta,q)
\end{eqnarray}
\end{enumerate}

\section{Numerical results and discussion}

\section*{Statistical properties}

\begin{figure}[H]	
 \centering
	\includegraphics[width=7.5cm, height=6.5cm]{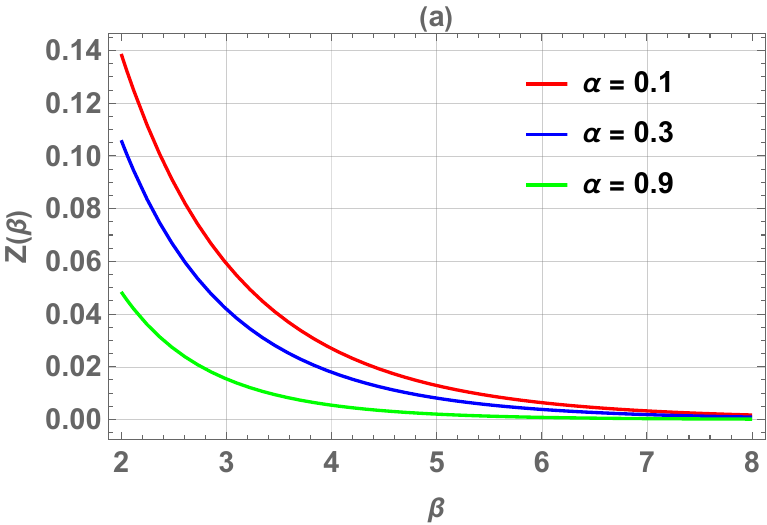}\quad
\includegraphics[width=7.5cm, height=6.5cm]{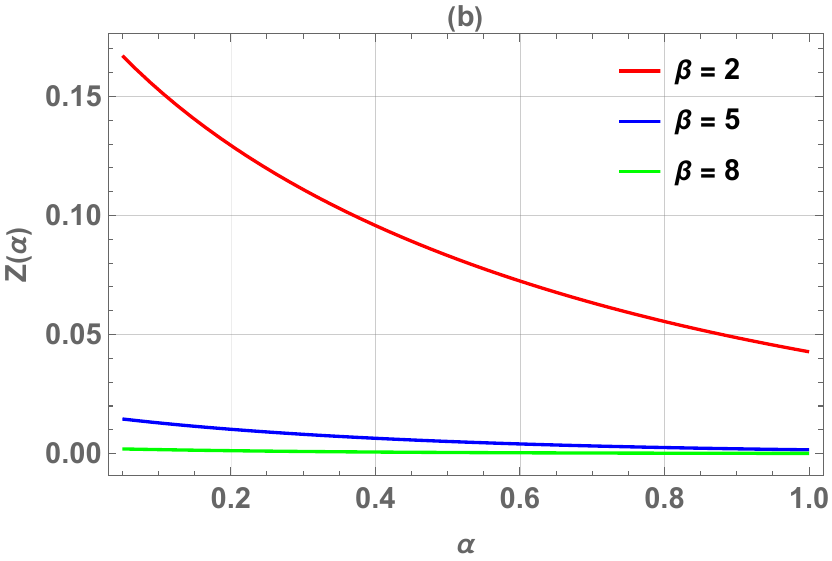}
\caption{ (The left panel(a), the partition function Z is displayed as a function of Inverse temperature parameter $\beta$, for various and fixed values of $\alpha$, and . (The right panel(b), the partition function Z is displayed as a 
function of $\alpha$, for various values of parameter $\beta$ which take tree valuers $2,5$ and $8$}.\label{Zen}
\end{figure}

The figures Fig.\ref{Zen}(a) and Fig.\ref{Zen}(b) investigates the behavior of the partition function \( Z \), a central quantity in statistical mechanics that encapsulates the thermodynamic properties of the system. In Fig.\ref{Zen}(a), \( Z(\beta) \) is plotted as a function of inverse temperature \( \beta \) for three distinct values of the interaction parameter \( \alpha \) (0.1, 0.3, and 0.9). The partition function exhibits a monotonically decreasing trend with increasing \( \beta \), reflecting the reduction in accessible microstates as the system approaches lower temperatures. Additionally, higher values of \( \alpha \) lead to significantly lower values of \( Z \), suggesting that the interaction term governed by \( \alpha \) suppresses the system’s configurational entropy and reduces the number of energetically favorable states. In Fig.\ref{Zen}(b), the focus shifts to the behavior of \( Z \) as a function of \( \alpha \), for fixed values of \( \beta = 2, 5, \) and \( 8 \). Here, \( Z(\alpha) \) is a decreasing function, with steeper declines observed for larger \( \beta \), i.e., lower temperatures. This reinforces the interpretation that the parameter \( \alpha \) acts to constrain the phase space, particularly at lower thermal energies. The observed smoothness and regularity of the partition function across both figures suggest a well-defined thermodynamic regime without singularities, consistent with non-critical systems. Overall, these plots underscore the fundamental role of \( \alpha \) in shaping the statistical weight of accessible states and demonstrate the interplay between thermal excitation and interaction strength in governing the equilibrium behavior of the system.

\begin{figure}[H]	
 \centering
	\includegraphics[width=7.5cm, height=6.0cm]{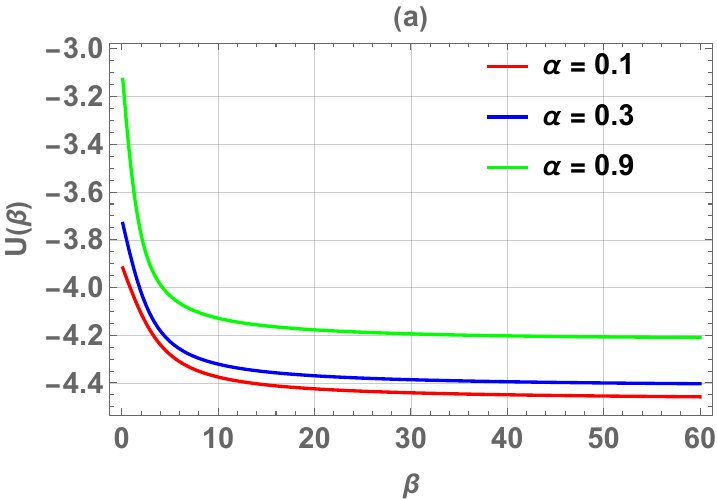}\quad
		\includegraphics[width=7.5cm, height=6.0cm]{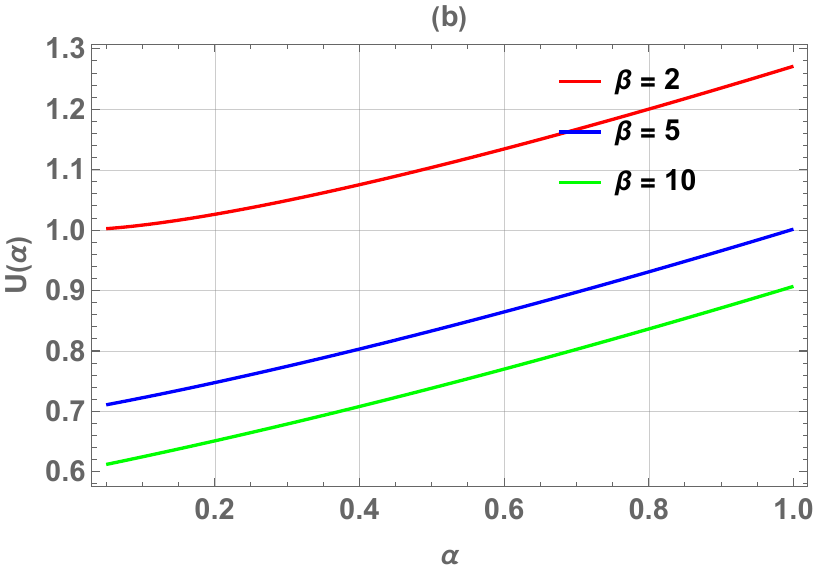}
\caption{: (The left panel(a), the mean energy U is displayed as a function of for various 
values of Inverse temperature parameter $\beta$, and . (The right panel(b), the mean energy is displayed as a function of $\alpha$, for various values of deformed parameter $\beta$.
}\label{UU}
\end{figure}

The Fig.\ref{UU} provide insight into the thermal evolution of the average energy \( U \), a key thermodynamic observable, as a function of inverse temperature \( \beta \) and interaction parameter \( \alpha \). In Fig.\ref{UU}(a), \( U(\beta) \) is shown for three values of \( \alpha \): 0.1, 0.3, and 0.9. The energy decreases monotonically with increasing \( \beta \), which corresponds to decreasing temperature, indicating that the system tends toward lower energy states as thermal agitation diminishes. For all values of \( \beta \), higher values of \( \alpha \) consistently produce lower energies, reflecting the stabilizing influence of this parameter and suggesting that it enhances the tendency of the system to occupy energetically favorable configurations. In contrast, Fig.\ref{UU}(b) examines \( U(\alpha) \) for fixed values of \( \beta \) (2, 5, and 10). Here, the average energy increases with \( \alpha \), which might at first seem contradictory; however, this likely arises from a different scaling or parameter dependence in this specific model. At higher \( \beta \) (lower temperatures), the energy remains lower overall but still shows a rising trend with increasing \( \alpha \). This implies that although \( \alpha \) generally acts as a stabilizing factor when the temperature varies, it also influences the energy landscape in a non-trivial way when held against a fixed thermal background. These results highlight the multifaceted role of \( \alpha \) in the thermodynamic behavior of the system, which governs both the depth and the accessibility of the energy minima. The smooth and monotonic nature of both plots further reinforces the picture of a well-behaved equilibrium system with no apparent phase transitions within the explored parameter range.

\begin{figure}[H]	
 \centering
	\includegraphics[width=7.5cm, height=6cm]{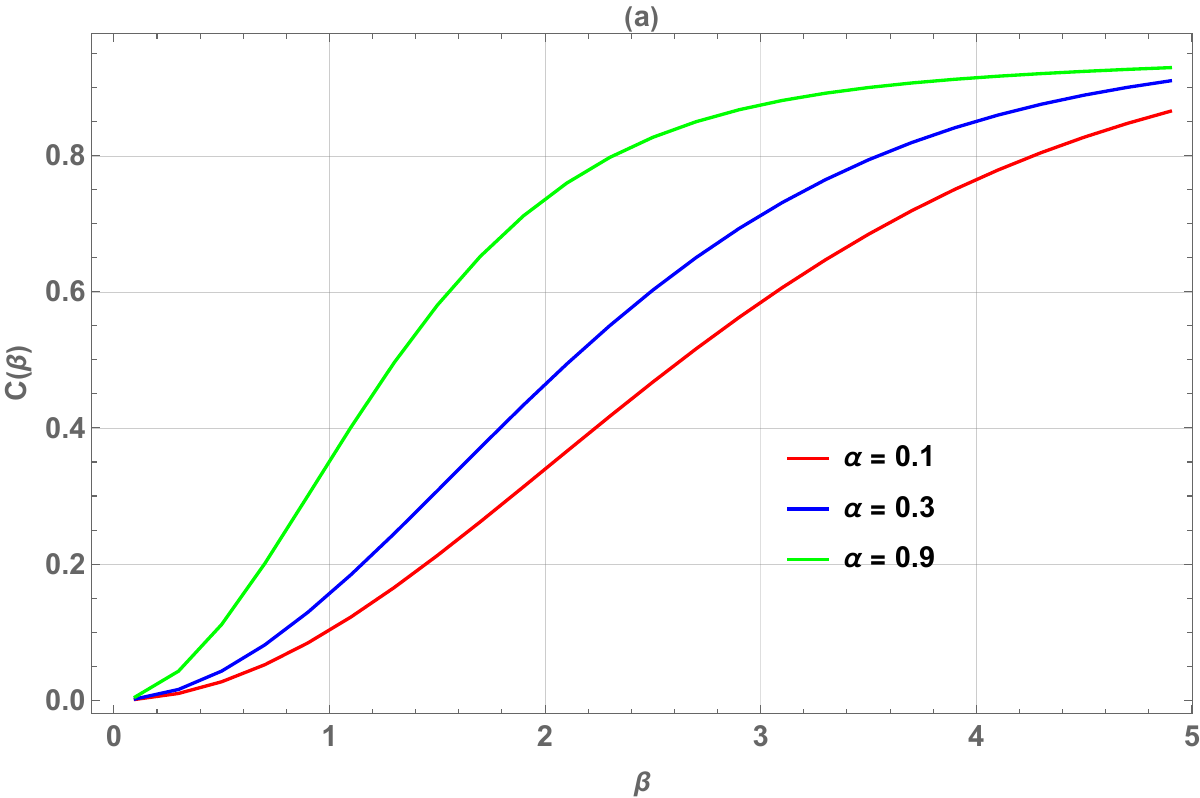}\quad
		\includegraphics[width=7.5cm, height=6cm]{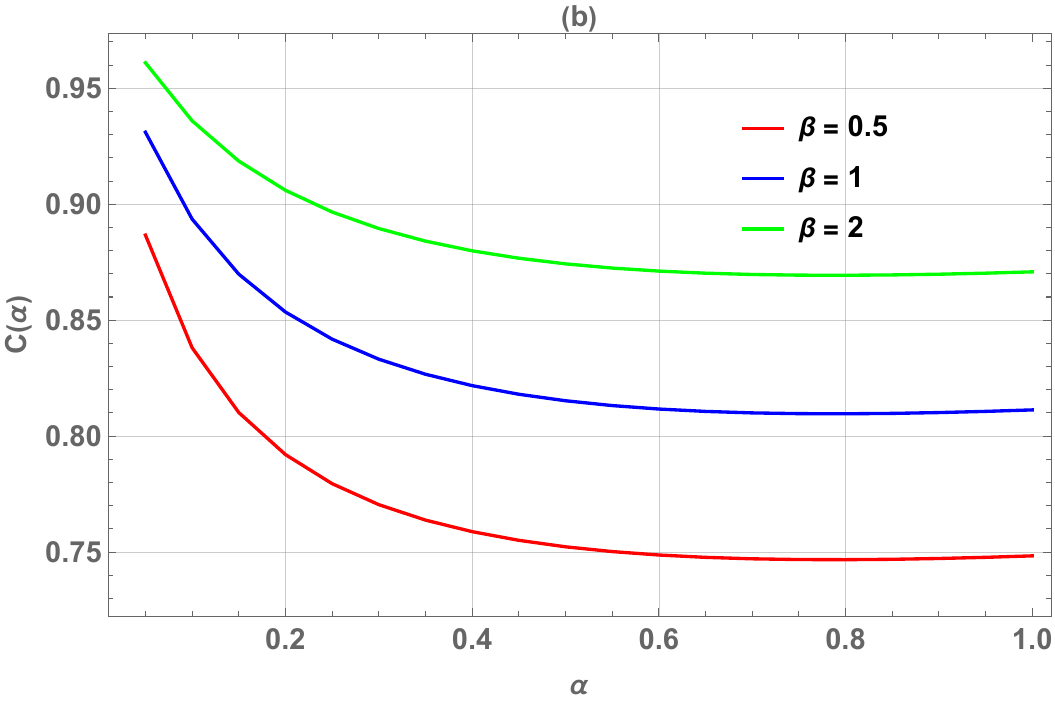}
\caption{ (The left panel(a), the specific heat C is displayed as a function of for various 
values of Inverse temperature parameter $\beta$, and  (The right panel(b), is displayed as a function of $C$, for various values of $\alpha$ for fixed deformed parameter $\beta$.
}\label{CC}
\end{figure}

The Fig\ref{CC} provide complementary views of the system’s heat capacity \( C \), highlighting its dependence on the inverse temperature \( \beta \) and the interaction parameter \( \alpha \). In Fig.\ref{CC}(a), \( C(\beta) \) is plotted for three values of \( \alpha \): 0.1, 0.3, and 0.9. The heat capacity increases sharply at first with \( \beta \), suggesting a rapid enhancement of thermal sensitivity at lower temperatures, before reaching a saturation regime where additional increases in \( \beta \) yield diminishing returns in \( C \). This behavior is indicative of a system transitioning from a disordered high-temperature regime to a more rigid low-temperature phase where thermal excitations become increasingly suppressed. As expected, larger values of \( \alpha \) result in reduced heat capacity, implying that stronger interactions (or coupling) constrain the degrees of freedom available for thermal energy storage. In Fig.\ref{CC}(b), the heat capacity is examined as a function of \( \alpha \), for three fixed values of \( \beta \): 0.5, 1, and 2. In this representation, \( C(\alpha) \) decreases monotonically with increasing \( \alpha \), again confirming that greater coupling leads to a stiffer, less thermally responsive system. Furthermore, the influence of \( \beta \) is evident: lower values of \( \beta \) (i.e., higher temperatures) correspond to higher capacities, as thermal fluctuations dominate. Together, the two figures portray a consistent thermodynamic narrative where the heat capacity reflects the interplay between thermal agitation and interaction-induced rigidity. The smooth and monotonic trends in both plots suggest a lack of critical phenomena and instead support a continuously varying, well-behaved response typical of analytically tractable models.

\begin{figure}[H]	
 \centering
	\includegraphics[width=7.5cm, height=6.5cm]{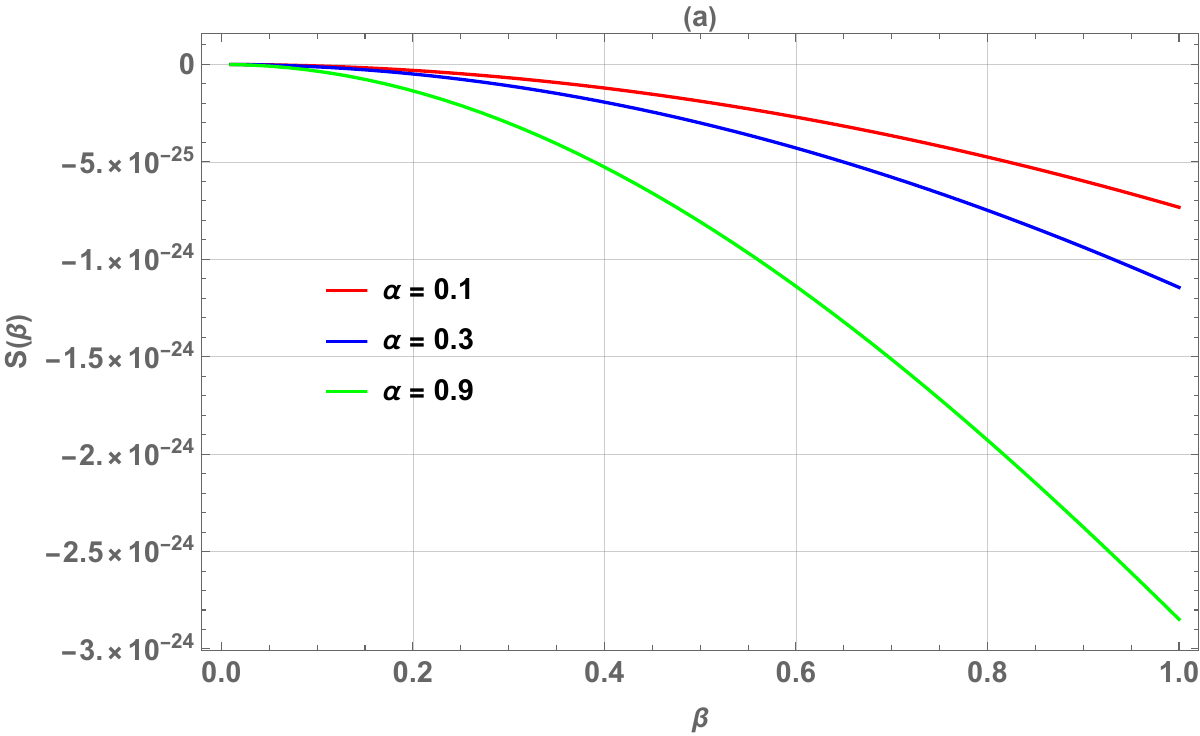}\quad
	\includegraphics[width=7.5cm, height=6.5cm]{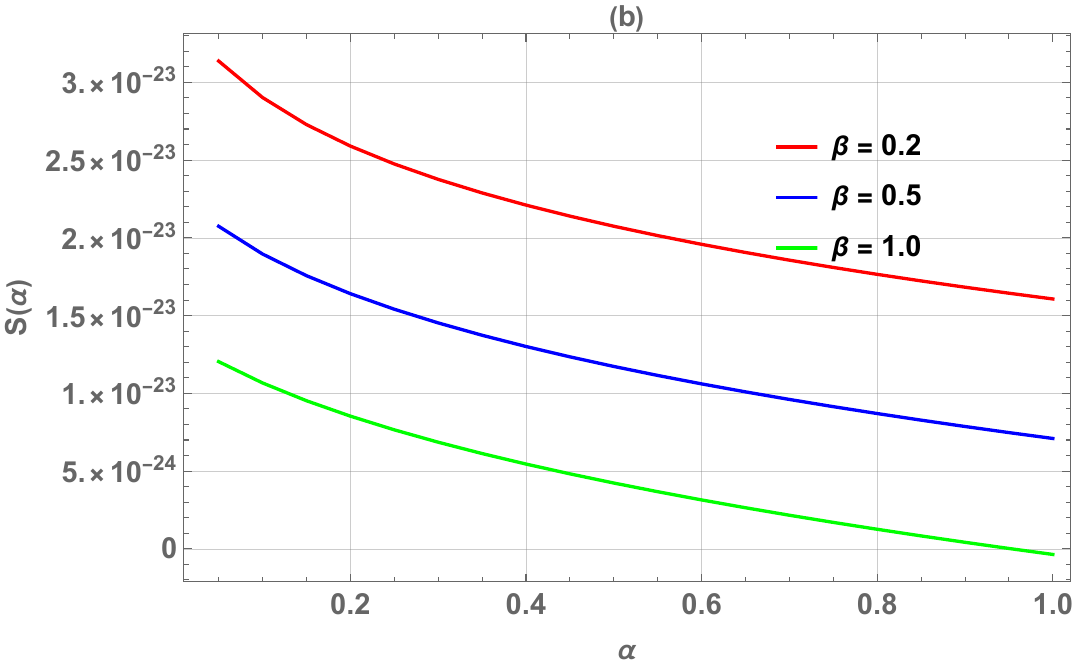}
\caption{: (The left pane(a)), the entropy S is displayed as a function of Inverse temperature parameter $\beta$ for various values 
of $\alpha$, and  (The right panel(b), is displayed as a function of $\alpha$, for various values of 
deformed parameter $\beta$.
}\label{SS}
\end{figure}

The two Fig.\ref{SS} depict the thermodynamic behavior of entropy \( S \) as a function of inverse temperature \( \beta \) and the interaction parameter \( \alpha \), respectively. In Fig.\ref{SS}(a), \( S(\beta) \) is shown for three values of \( \alpha \): 0.1, 0.3, and 0.9. The entropy decreases monotonically as \( \beta \) increases, consistent with the fundamental thermodynamic principle that entropy diminishes at lower temperatures due to a reduction in accessible microstates. Furthermore, for any fixed \( \beta \), increasing \( \alpha \) results in lower entropy values, indicating that the interaction parameter \( \alpha \) effectively constrains the configurational degrees of freedom of the system. The entropy values are notably small and negative, on the order of \( 10^{-24} \), which suggests the use of a relative entropy scale or a renormalized reference state. Fig.\ref{SS}(b), on the other hand, presents \( S(\alpha) \) for fixed values of \( \beta = 0.2, 0.5, \) and \( 1.0 \). In this case, entropy increases significantly with \( \alpha \), particularly at lower \( \beta \) (i.e., higher temperatures), reflecting a regime where interaction strength enhances rather than suppresses entropy. This seemingly contrasting trend can be interpreted as a reflection of the dual influence of \( \alpha \) and \( \beta \): at high thermal energies, increasing \( \alpha \) may broaden the distribution of accessible states, thus increasing entropy. Overall, the results suggest that the entropy's dependence on \( \alpha \) is temperature-sensitive and non-monotonic across thermal regimes. The smooth evolution and absence of singular behavior in both figures affirm the analytic nature of the underlying model and its suitability for controlled thermodynamic analysis.

\begin{figure}[H]
 \centering
	\includegraphics[width=7.5cm, height=6.5cm]{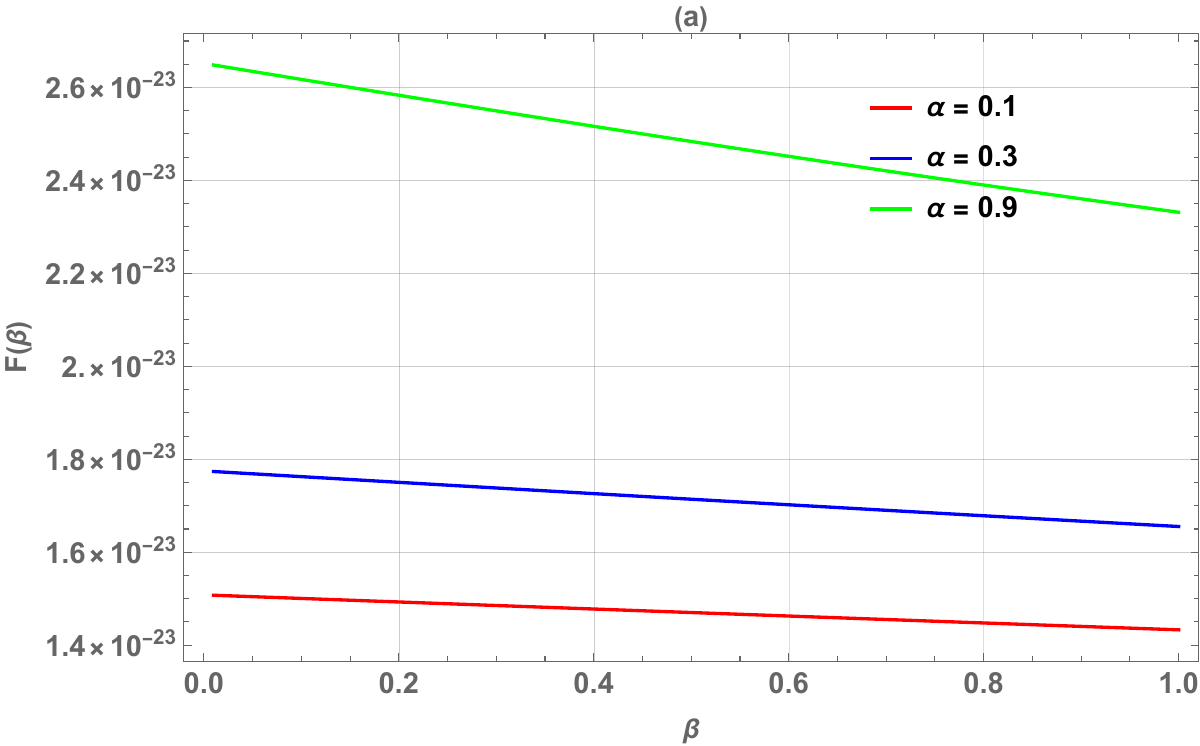}\quad
	\includegraphics[width=7.5cm, height=6.5cm]{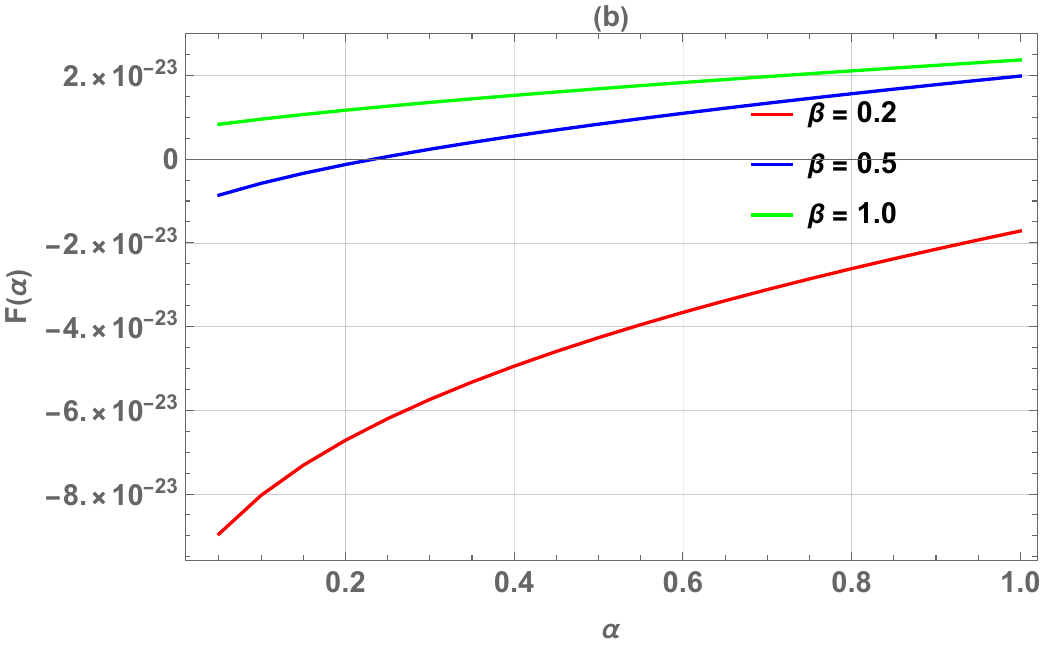}
\caption{: (The left panel(a), the free energy F is displayed for various values of $\alpha$ as a 
function of Inverse temperature parameter $\beta$, and  (The right panel(b), is displayed as a function of $\alpha$, for various values 
of deformed parameter $\beta$.
}\label{FF}
\end{figure}

The two Fig.\ref{FF} offer a thermodynamic analysis of the Helmholtz free energy \( F \), plotted respectively as a function of inverse temperature \( \beta \) and the interaction parameter \( \alpha \). In Fig.\ref{FF}(a), \( F(\beta) \) is illustrated for three values of \( \alpha \): 0.1, 0.3, and 0.9. The free energy increases monotonically with \( \beta \), which corresponds to decreasing temperature, reflecting the growing contribution of internal energy as thermal disorder subsides. For any fixed \( \beta \), larger values of \( \alpha \) yield lower values of \( F \), indicating a stabilizing influence of this interaction parameter that reduces the system's overall thermodynamic potential. The behavior is smooth and nearly linear, suggesting an absence of abrupt transitions. In Fig.\ref{FF}(b), \( F(\alpha) \) is shown for fixed values of \( \beta = 0.2, 0.5, \) and \( 1.0 \). In contrast to Fig.\ref{FF}(a), the free energy here decreases as \( \alpha \) increases, with curves exhibiting a quasi-linear descent whose steepness depends on the value of \( \beta \). For lower \( \beta \) (higher temperatures), the free energy remains positive, while at higher \( \beta \), \( F \) becomes negative for larger \( \alpha \), signifying a stronger stabilizing effect at lower temperatures. This dual representation reveals how the interplay between thermal energy and interaction strength governs the system’s equilibrium configuration. The consistent monotonicity and smooth gradients in both graphs affirm the analytic and stable character of the model, indicative of a system that remains far from critical points within the explored parameter space.

\section*{Superstatistic properties}

\begin{figure}[H]	
 \centering
	\includegraphics[width=7.5cm, height=6.5cm]{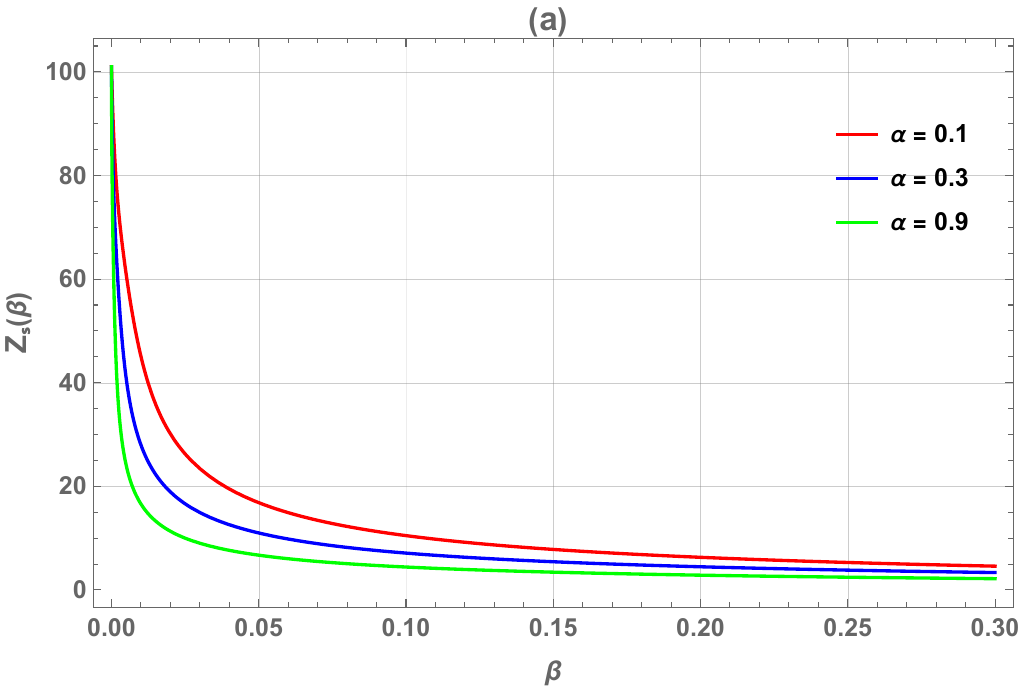}\quad
		\includegraphics[width=7.5cm, height=6.5cm]{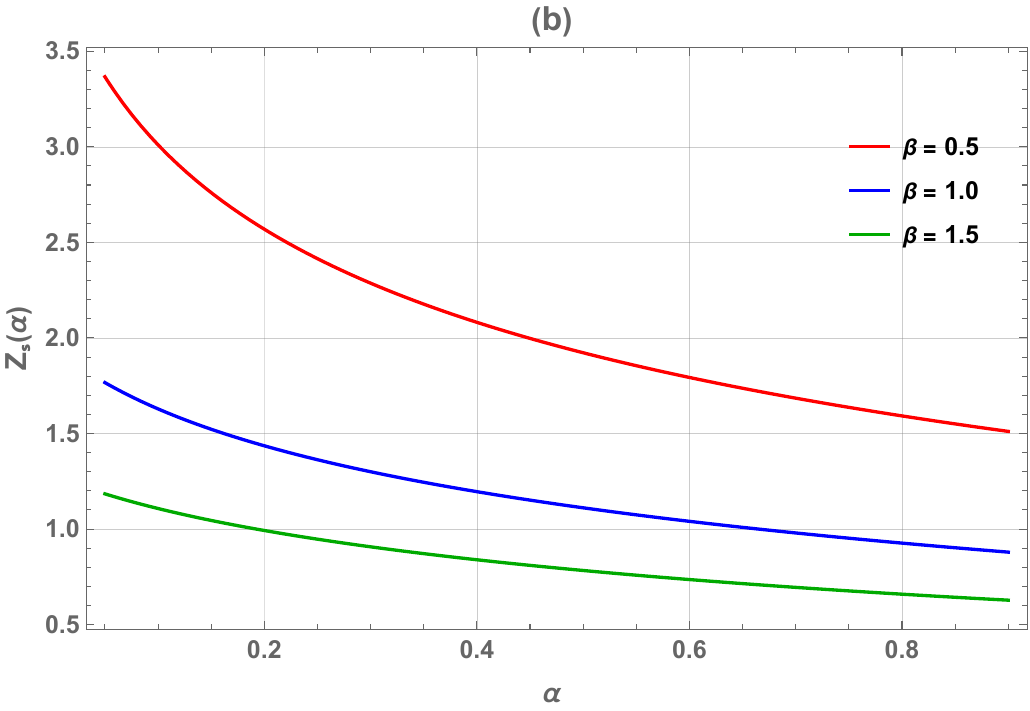}
\caption{ (The left panel(a), the superstatistical partition function $Z_s$ is displayed as a function of Inverse temperature parameter $\beta$, for various and fixed values of $\alpha$, and . (The right panel(b), this partition function $Z_s$ is displayed as a 
function of $\alpha$, for various values of  parameter $\beta$ which take tree valuers $0.5,1.0$ and $1.5$}\label{Zs}
\end{figure}

Figures Fig.\ref{Zs}(a) and (b) illustrate the behavior of the supertatistic partition function \( Z_s \) as a function of the inverse temperature \( \beta \) an the deformation parameter \( \alpha \) respectively. In Fig.\ref{Zs}(b), \( Z_s \) decreases monotonically with increasing \( \alpha \), indicating that mass deformation reduces the number of accessible microstates by introducing geometric constraints. This suppression becomes more pronounced at higher \( \beta \) (i.e., lower temperatures). Fig.\ref{Zs}(a)  shows that \( Z_s \) increases with \( \beta \), consistent with the thermal occupation of low-energy states, but this growth is significantly dampened for larger \( \alpha \). Overall, the results highlight that \( \alpha \) acts as a superstatistical parameter that modulates thermal accessibility and energy-level sensitivity of the system.

\begin{figure}[H]	
 \centering
	\includegraphics[width=7.5cm, height=6.5cm]{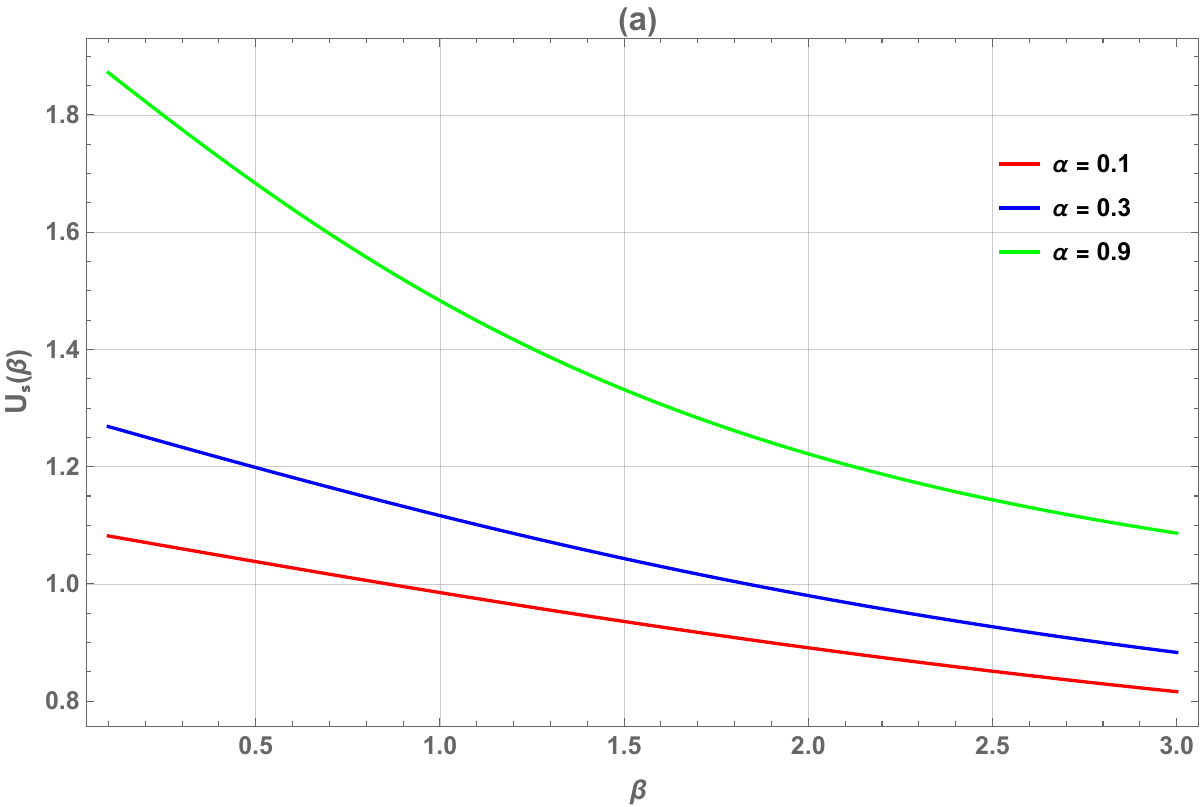}\quad
	\includegraphics[width=7.5cm, height=6.5cm]{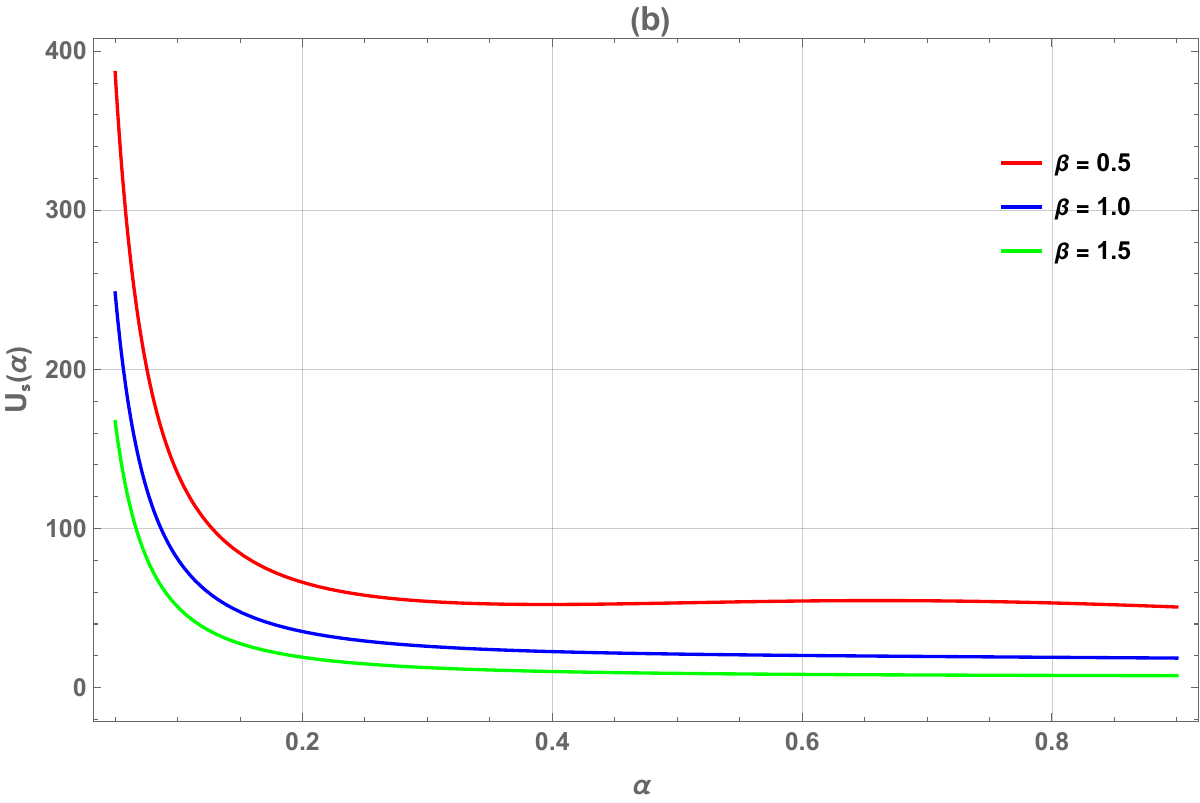}
\caption{: (The left panel(a), the superstatistic mean energy $U_s$ is displayed as a function of for various 
values of Inverse temperature parameter $\beta$, and . (The right panel(b) in a function of $\alpha$, for various values of deformed parameter $\beta$.
}\label{Us}
\end{figure}

Fig.\ref{Us}(a) and (b) present the behavior of the superstatistical internal energy \( U_s \) as a function of the inverse temperature \( \beta \) and the deformation parameter \( \alpha \), respectively. In Fig.\ref{Us}(a), \( U_s(\beta) \) decreases monotonically as \( \beta \) increases, consistent with the thermal depopulation of higher energy states at lower temperatures. The magnitude of this decrease is more pronounced for lower \( \alpha \), showing that less deformed systems are more thermally responsive. In Fig.\ref{Us}(b), \( U_s(\alpha) \) increases rapidly with \( \alpha \), especially at higher \( \beta \), indicating that mass deformation leads to a steeper energy spectrum and higher excitation energy. This reveals that \( \alpha \) enhances confinement and raises the overall energy cost to populate quantum states. These trends confirm the role of \( \alpha \) as a structural control parameter influencing the thermodynamic rigidity of the system.

\begin{figure}[H]	
 \centering
		\includegraphics[width=7.5cm, height=6.5cm]{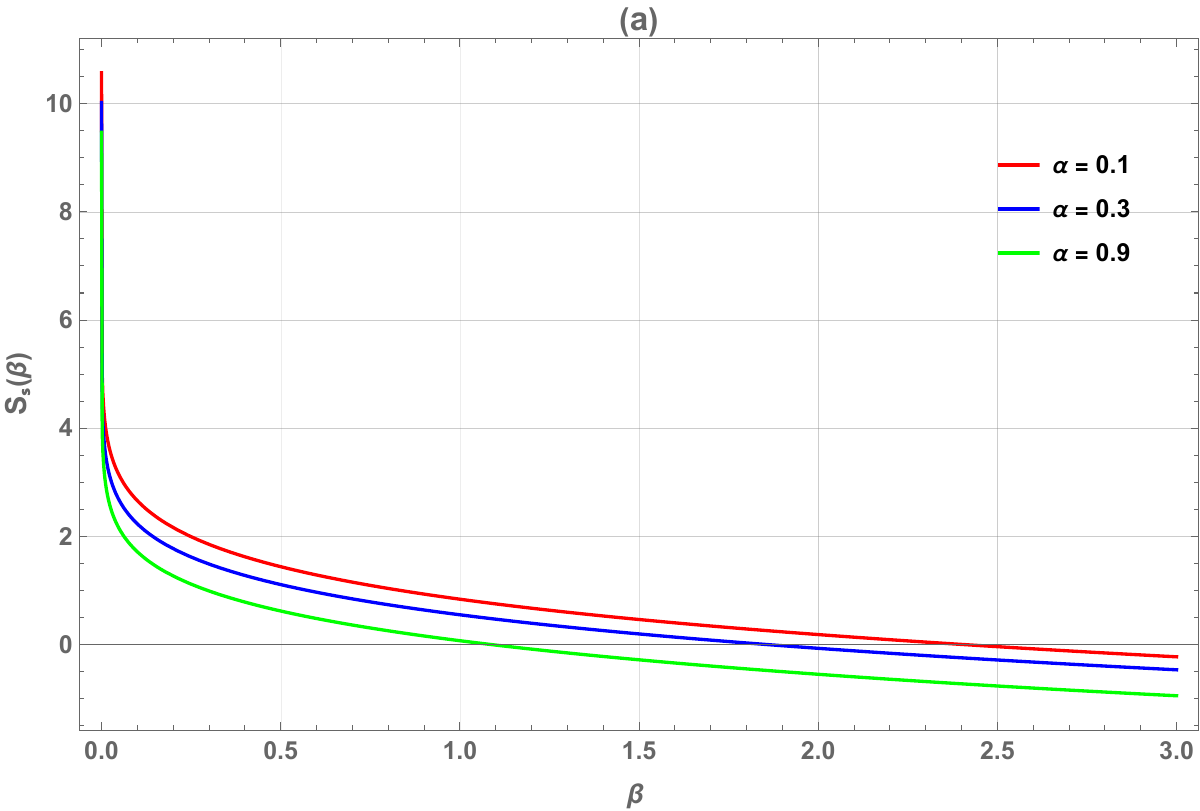}\quad
	\includegraphics[width=7.5cm, height=6.5cm]{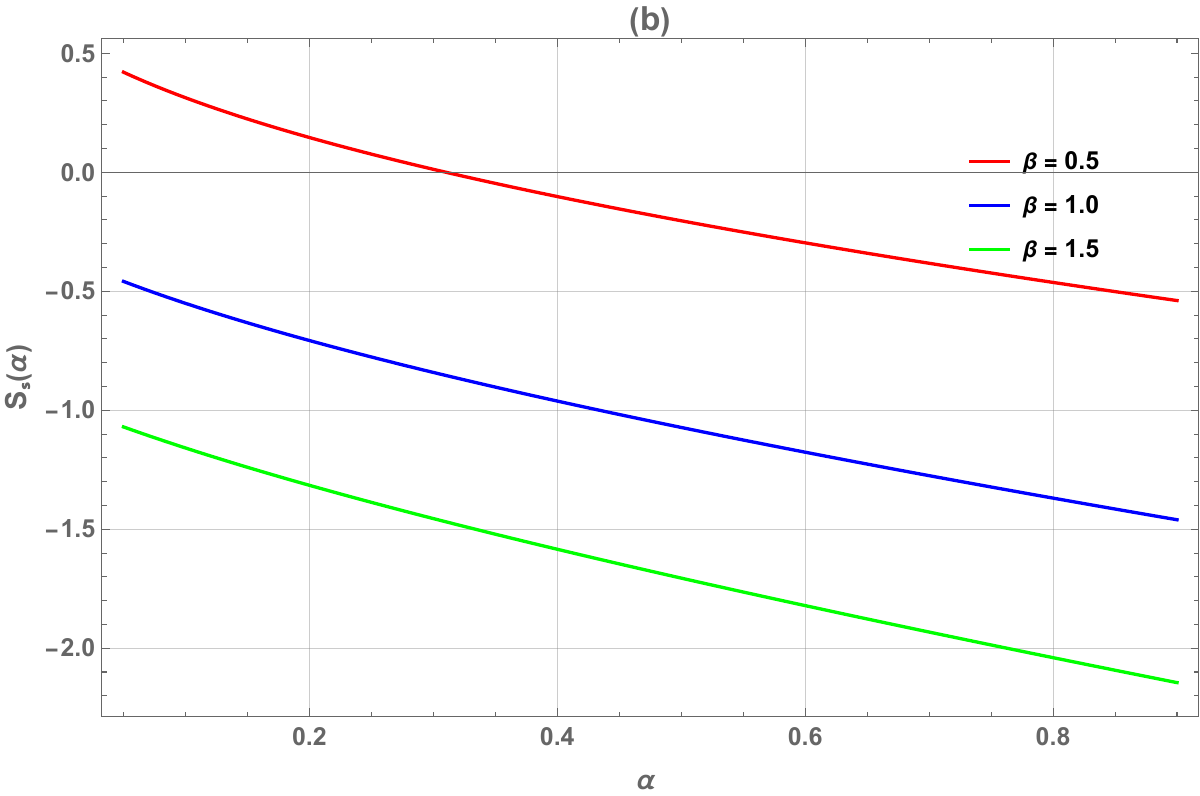}\label{S}
\caption{: (The left pane(a)), the superstatistic entropy S is displayed as a function of Inverse temperature parameter $\beta$ for various values 
of $\alpha$, and  (The right panel(b), is displayed as a function of $\alpha$, for various values of 
deformed parameter $\beta$.
}\label{Ss}
\end{figure}

The Fig.\ref{Ss}(a) and (b) illustrate the dependence of the superstatistic entropy \( S_s \) on the inverse temperature \( \beta \) and the deformation parameter \( \alpha \), respectively. In Figure (a), \( S_s \) increases with \( \beta \), reflecting the enhanced thermal population of energy levels and greater statistical disorder at lower temperatures. However, systems with higher \( \alpha \) exhibit consistently lower entropy, suggesting that deformation suppresses the number of accessible microstates. Figure (b) shows that \( S_s \) decreases steadily as \( \alpha \) increases, with this decline becoming sharper at higher values of \( \beta \). This behavior indicates that mass deformation acts as an ordering mechanism, restricting phase space and limiting thermodynamic fluctuations. Overall, the results emphasize the role of \( \alpha \) as a controlling parameter that reduces entropy, especially in low-temperature regimes.

\begin{figure}[H]
 \centering
	\includegraphics[width=7.5cm, height=6.5cm]{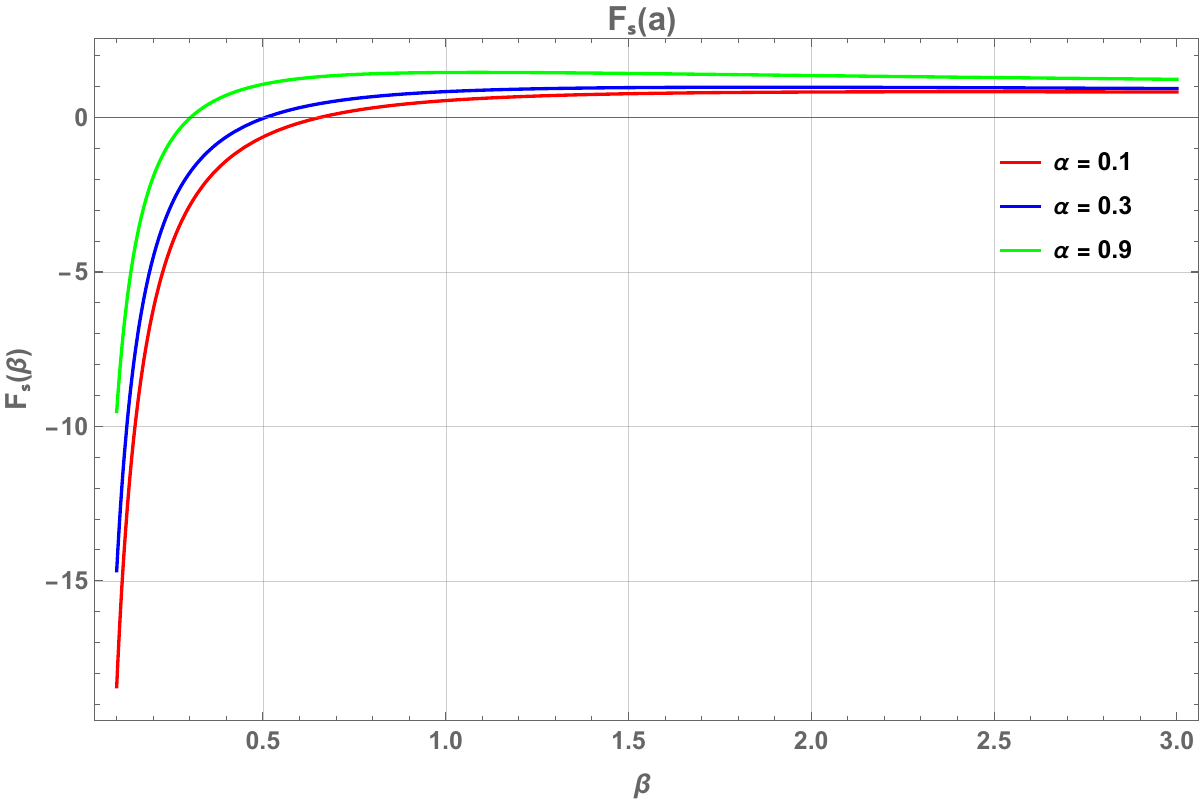}\quad
	\includegraphics[width=7.5cm, height=6.5cm]{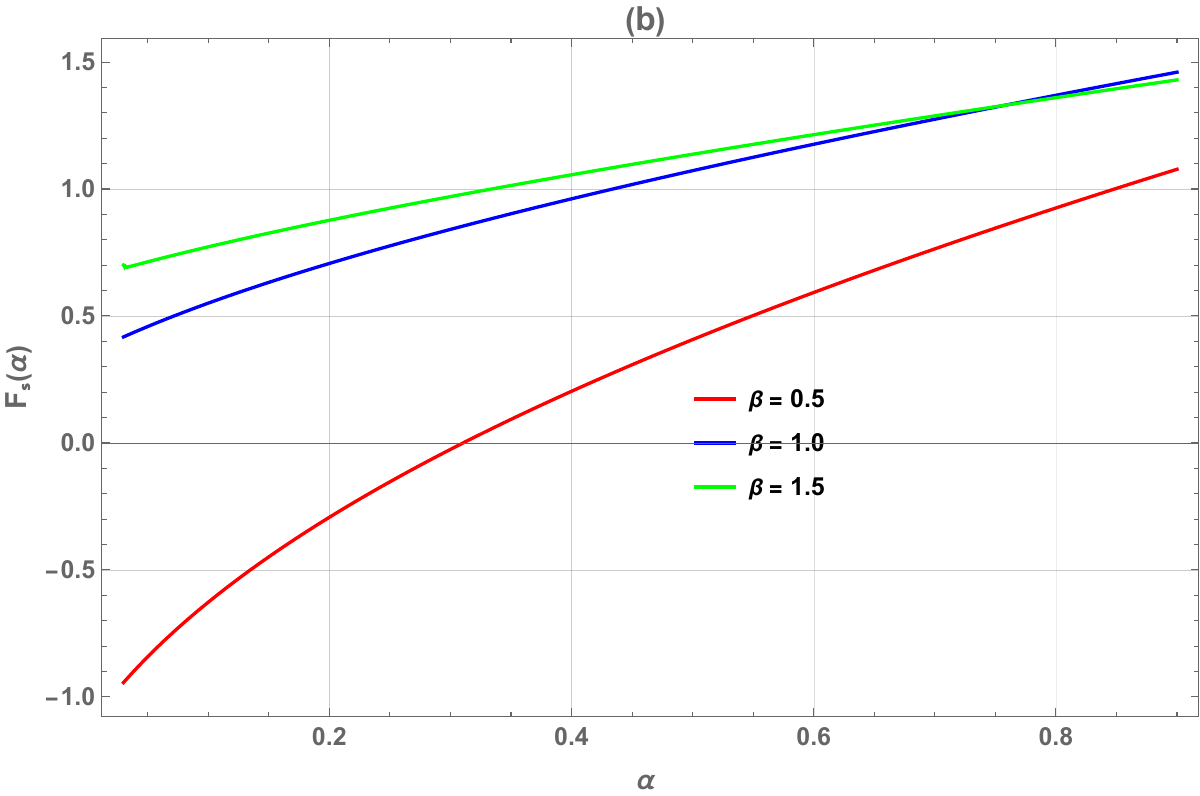}\label{F}
\caption{: (The left panel(a), the superstatistic free energy F is displayed for various values of $\alpha$ as a 
function of Inverse temperature parameter $\beta$, and  (The right panel(b), is displayed as a function of $\alpha$, for various values 
of deformed parameter $\beta$.
}\label{Fs}
\end{figure}

In the Fig.\ref{Fs}(a) and (b) display the variation of the superstatistic Helmholtz free energy \( F_s \) with respect to the inverse temperature \( \beta \) and the deformation parameter \( \alpha \). In Fig\ref{Fs}(a), \( F_s \) decreases nonlinearly as \( \beta \) increases, becoming more negative, which is consistent with the system approaching its ground state energy at low temperatures. The rate of this decline is greater for lower values of \( \alpha \), indicating a stronger thermodynamic response in less deformed systems. In Fig.\ref{Fs}(b), \( F_s(\alpha) \) increases smoothly with \( \alpha \), particularly at higher \( \beta \), showing that deformation tends to reduce the thermodynamic cost of maintaining equilibrium. The results highlight that deformation not only modifies the energy spectrum but also stabilizes the system thermodynamically. Overall, \( \alpha \) softens the temperature dependence of free energy, acting as a structural regulator of equilibrium properties.

\begin{figure}[H]	
 \centering
	\includegraphics[width=7.5cm, height=6cm]{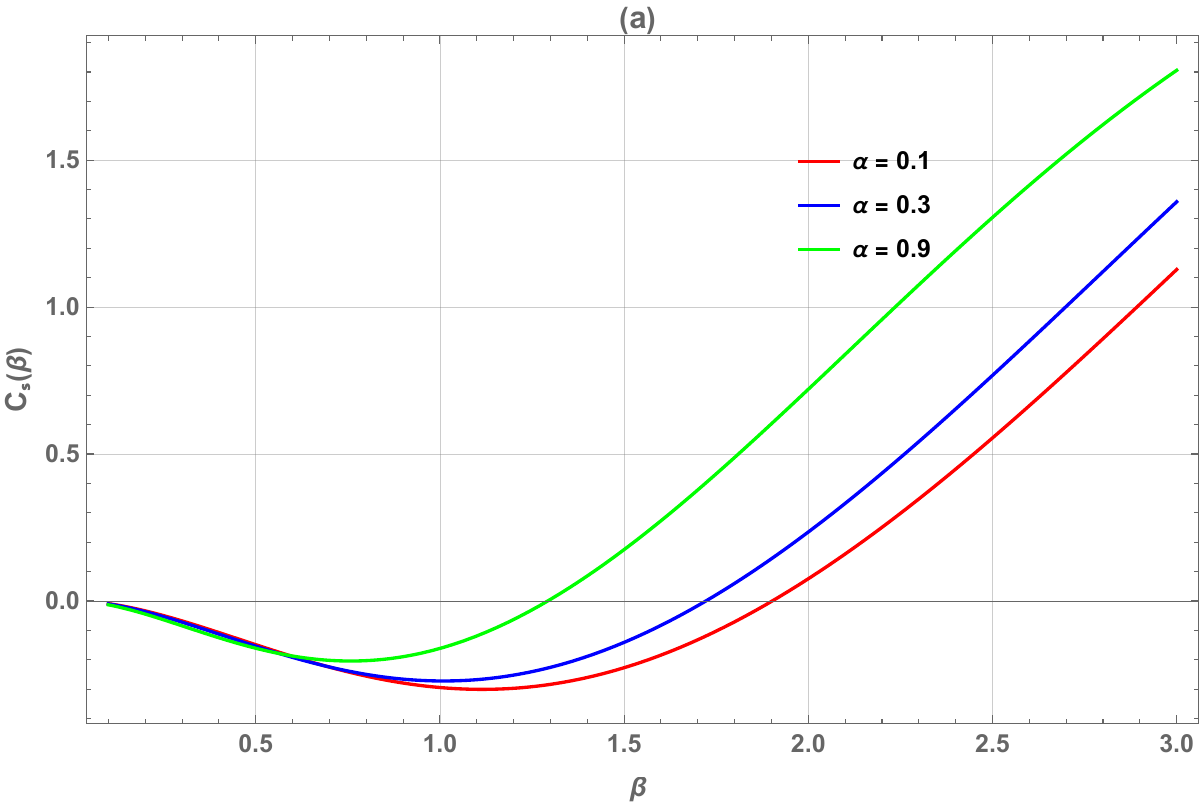}\quad
	\includegraphics[width=7.5cm, height=6cm]{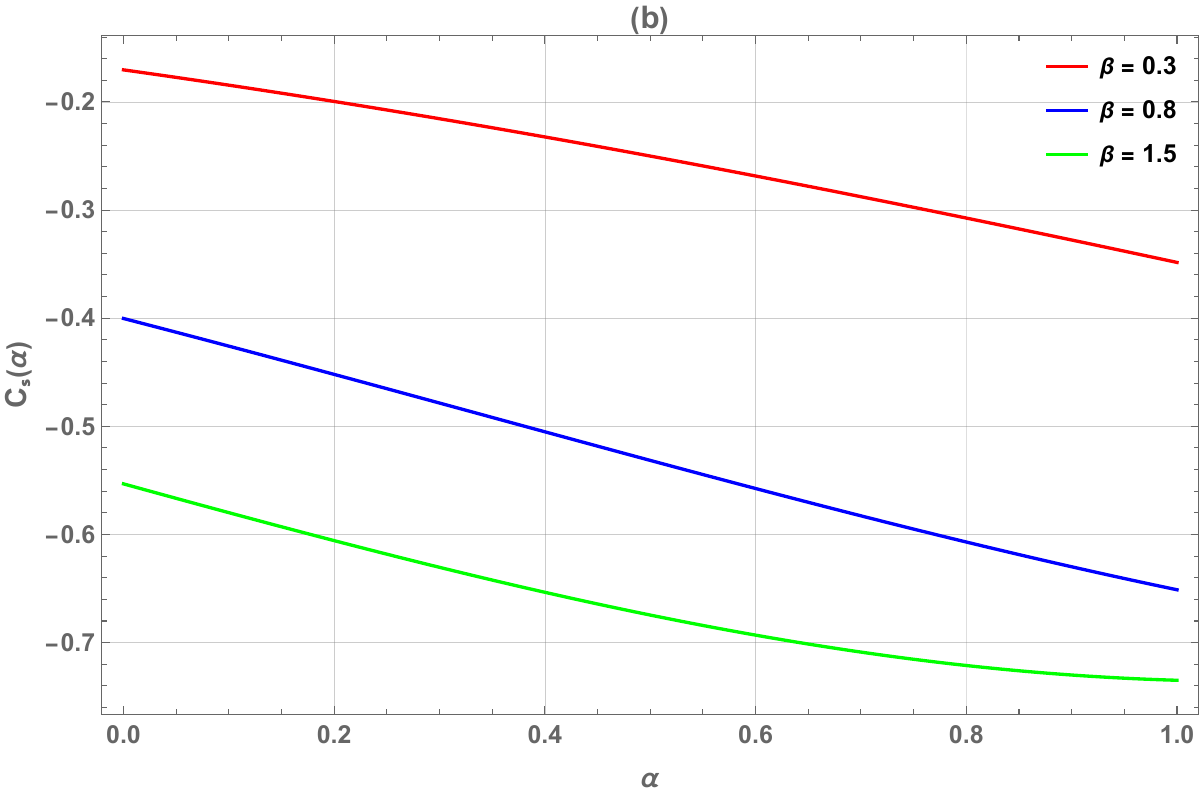}
\caption{ (The left panel(a), the superstatistic specific heat $C_s$ is displayed as a function of for various 
values of Inverse temperature parameter $\beta$, and  (The right panel(b), is displayed as a function of $C_s$, for various values of $\alpha$ for fixed deformed parameter $\beta$.
}\label{Cs}
\end{figure}

In the Fig.\ref{Cs}(a) , $C_s(\beta)$ shows a smooth increase with the inverse temperature parameter $\beta$, saturating at large $\beta$, which is consistent with thermodynamic behavior under superstatistical regimes. This trend indicates enhanced energy fluctuations at intermediate $\beta$, often associated with metastable or transition regions. In contrast, Fig.\ref{Cs}(b) shows that $C_s(\alpha)$ decreases monotonically with increasing position-dependent mass parameter $\alpha$, for all fixed values of $\beta$. This behavior suggests that greater spatial variation in mass inhibits thermal excitations, thus lowering the specific heat. These observations are consistent with prior studies on systems governed by position-dependent mass and generalized coherent states, where both temperature and mass modulation significantly influence thermal response and statistical properties.

\section{Conclusion}\label{sec4}

This work explores the thermodynamic and superstatistical properties of a harmonic oscillator with a position-dependent mass (PDM) under a quadratic deformation governed by a parameter $\alpha$. By solving the deformed Schrödinger equation analytically, the authors derive energy spectra and Gazeau–Klauder coherent states, which are then used to compute thermodynamic quantities such as partition function $Z(\beta)$, mean energy  $U(\beta)$, entropy $S(\beta)$, heat capacity 
 $C(\beta)$, and free energy  $F(\beta)$. The numerical analysis reveals that the deformation parameter $\alpha$ plays a key role in modulating thermal behavior: increasing $\alpha$ reduces entropy and specific heat, indicating stronger confinement and fewer accessible microstates. The results also show smooth, monotonic trends in all thermodynamic functions, suggesting the absence of critical behavior. Moreover, the superstatistical framework enhances the understanding of thermal fluctuations in systems with inhomogeneous structures, showing that $\alpha$ can be used as a tunable parameter to control quantum thermodynamic responses. These insights contribute meaningfully to the theoretical modeling of nanoscale systems and coherent quantum states in deformed environments.

\section*{Data Availability Statement }
This manuscript has associated data in a data repository. [Authors’ comment: All 
data included in this manuscript are available 
on request by contacting the corresponding author]. 
\section*{Declaration of competing interest} 
 The authors declare that they have no known competing 
financial interests or personal relationships that could have appeared to influence the  work reported in this paper. 

\section*{Data availability}
 No data was used for the research described in the article.

\section*{Acknowledgments}
We acknowledges LMT for useful discussions.

\end{document}